\documentclass[conference]{IEEEtran}
\IEEEoverridecommandlockouts
\usepackage{cite}
\usepackage{amsmath,amssymb,amsfonts}
\usepackage{algorithmic}
\usepackage{graphicx}
\graphicspath{{graphs/}} 
\usepackage{textcomp}
\usepackage{xcolor}
\usepackage{soul}
\usepackage{url}
\usepackage[hidelinks]{hyperref}
\usepackage{orcidlink}
\usepackage{caption}
\usepackage{subcaption}
\usepackage{booktabs}
\newif\ifshowchanges
\showchangestrue

\begin{document}
\pagenumbering{gobble}

\title{Fine-Grained Power and Energy Attribution on AMD GPU/APU-Based Exascale Nodes\\ 
\thanks{Notice: This manuscript has been authored by UT-Battelle LLC under contract DE-AC05-00OR22725 with the US Department of Energy (DOE). The US government retains and the publisher, by accepting the article for publication, acknowledges that the US government retains a nonexclusive, paid-up, irrevocable, worldwide license to publish or reproduce the published form of this manuscript, or allow others to do so, for US government purposes. DOE will provide public access to these results of federally sponsored research in accordance with the DOE Public Access Plan (\url{https://www.energy.gov/doe-public-access-plan}).}
}

 \author{%
\IEEEauthorblockN{%
 Adam McDaniel\,\orcidlink{0000-0001-6926-028X}\IEEEauthorrefmark{1},
 Michael Jantz\,\orcidlink{0000-0003-4460-1206}\IEEEauthorrefmark{1},
 Ashesh Sharma\,\orcidlink{0009-0005-9959-050X}\IEEEauthorrefmark{2},
 Steve Abbott\,\orcidlink{0009-0000-1028-819X}\IEEEauthorrefmark{2}
 Steven Martin\,\orcidlink{0000-0002-8489-2012}\IEEEauthorrefmark{2},
 Shreyas Khandekar\,\orcidlink{0009-0005-9351-6157}\IEEEauthorrefmark{2}, \\
 Brandon Neth\,\orcidlink{0009-0005-2692-8266}\IEEEauthorrefmark{2},
Bruno Villasenor Alvarez\,\orcidlink{0000-0002-7460-8129}\IEEEauthorrefmark{3},
Aditya Kashi \,\orcidlink{0000-0003-2589-3792}\IEEEauthorrefmark{4},
 Wael Elwasif\,\orcidlink{0000-0003-0554-1036}\IEEEauthorrefmark{4},
 Oscar Hernandez\,\orcidlink{0000-0002-5380-6951}\IEEEauthorrefmark{4}
 }

 \IEEEauthorblockA{\IEEEauthorrefmark{1}\textit{Dept. of Electrical Engineering and Computer Science, University of Tennessee}, Knoxville, TN, USA\\ amcdan23@vols.utk.edu,\; mrjantz@utk.edu}

\IEEEauthorblockA{\IEEEauthorrefmark{2}\textit{Hewlett Packard Enterprise (HPE)}, Bloomington, MN, USA\\
 ashesh.sharma@hpe.com,\; stephen.abbott@hpe.com,\; steven.martin3@hpe.com,\\\; shreyas.khandekar@hpe.com,\; brandon.neth@hpe.com}

 \IEEEauthorblockA{\IEEEauthorrefmark{3}\textit{Advanced Micro Devices, Inc.}, Santa Clara, CA, USA\\
 bruno.villasenoralvarez@amd.com}

 \IEEEauthorblockA{\IEEEauthorrefmark{4}\textit{Oak Ridge National Laboratory}, Oak Ridge, TN, USA\\
 kashia@ornl.gov,\; elwasifwr@ornl.gov,\; oscar@ornl.gov}
 }

\maketitle

\begin{abstract}
Modern exascale GPU- and APU-based systems provide multiple power and energy sensors, but differences in scope, update rate, timing, and filtering complicate the attribution of short-lived accelerator activity. This paper presents a methodology to characterize and correct these effects on Cray EX systems with AMD Instinct MI250X GPUs (Frontier) and MI300A APUs (Portage). Using controlled square-wave workloads, we quantify update intervals, delay, aliasing, and variability across up to 512 GPUs and 480 APUs with on-chip (\texttt{rocm-smi}/\texttt{amd-smi}) and off-chip Cray Power Management sensors. We reconstruct power from cumulative energy counters to achieve faster response times, validate it against on-chip, off-chip, and node-level sensors, and integrate the resulting streams into a Score-P/PAPI-based tool for time-aligned, phase-level attribution. Applied to rocHPL, rocHPL-MxP, and HPG-MxP, the method separates energy savings due to reduced runtime from changes in power. Mixed precision reduces node energy on Frontier by 79\% for rocHPL-MxP and 31\% for HPG-MxP, with similar trends on Portage. These results provide portable guidance for sensor validation and power-aware optimization on current and future exascale systems.

\end{abstract}


\section{Introduction}
Today’s supercomputers and large HPC data centers are increasingly constrained by power and thermal limits. The slowdown of Moore’s law, the rise of dense heterogeneous architectures, and the increasing external pressures on energy infrastructure amplify the need for energy-efficient, application-based optimizations. As power becomes a first-class design constraint, understanding how applications consume energy over time, across system components, and across fine-grained call paths is essential for identifying sources of inefficiency and guiding energy-aware optimizations on current systems.

Modern accelerator-based systems such as Frontier (with AMD\footnote{AMD, Instinct, the AMD Arrow logo and combinations thereof are trademarks of Advanced Micro Devices, Inc.} Instinct\texttrademark~MI250X GPUs) and Portage (with AMD Instinct\texttrademark~MI300A APUs, which have a similar architecture as El Capitan) expose a rich set of power and energy sensors through interfaces such as AMD's \texttt{rocm-smi} (which is being replaced by \texttt{amd-smi}), Cray Power Management (PM), and cabinet- or facility-level telemetry. These sensors vary widely in hardware scope, response times, update frequencies, averaging behavior, timestamp semantics, susceptibility to sample loss, and the overhead required for tools to read them. Built-in GPU counters can update at millisecond-scale resolution but may include undocumented filtering that affects responsiveness, while off-chip or node-level counters typically update more slowly and provide a coarser but independent view. This diversity complicates efforts to attribute power and energy to application activity, particularly under rapidly changing or bursty workloads.

To address this emerging complexity, this work develops a methodology for characterizing and correcting power and energy sensor behavior on heterogeneous nodes.
Our approach enables practitioners to make sense of measurements collected from disparate sources, including tools and platform-specific architectural counters.
While our present work is deployed only on cutting-edge AMD platforms, the primary contributions of this methodology are broadly applicable. These include: (a) a measurement model that separates sensor acquisition, driver publication, and tool sampling; (b) a confidence-window formalism for reliable steady-state attribution; (c) the use of $\Delta E / \Delta t$ to recover near-instantaneous power from cumulative energy counters; and (d) the use of synchronized traces to align heterogeneous sensor streams with application regions.

Accurate fine-grained power and energy attribution is important for application-level optimizations on modern heterogeneous systems. Interactions among short-lived kernels and rapidly changing phases can significantly affect power consumption, but are often obscured by sensor averaging and limited temporal resolution~\cite{10.1145/3703001.3724383,10.1007/978-3-032-07612-0_18}. Recovering these behaviors reveals phase boundaries and transient power excursions, enabling kernel-level power capping, scheduling, and library optimizations. Moreover, time-aligned traces with consistent device and region context support scalable data analytics for deeper performance and energy insights across nodes and system components.

We instantiate this methodology on two exascale-class systems that represent distinct integration models: Frontier, with discrete MI250X accelerators, and Portage, with integrated MI300A APUs. Using synthetic square-wave workloads, we characterize update intervals, delay, response, recovery, and aliasing, and then validate the approach on rocHPL, rocHPL-MxP, and HPG-MxP.
Hence, these experiments demonstrate measurement fidelity and phase-level attribution and under realistic application behavior.

Our contributions can be summarized as follows:
\begin{itemize}
  \item Measurement model and characterization method: We use square-wave workloads to evaluate sensor response time, delay, update rate, effective observation cadence, and aliasing on Frontier and Portage, and we formalize how these effects constrain fine-grained attribution.
  \item Portable attribution methodology: We correlate on-chip (\texttt{amd-smi} or \texttt{rocm-smi}) and off-chip (Cray PM) sensor data with application event timelines, derive high-fidelity instantaneous power from cumulative energy counters, and export synchronized traces that support scalable analytics for power and energy analysis.
  \item Application case studies: We apply the methodology to full- and mixed-precision HPL and HPG workloads, showing how phase-level attribution separates reduced time-to-solution from changes in instantaneous power and exposes actionable optimization opportunities.
\end{itemize}

By characterizing the behavior of these sensors and establishing confidence windows for attribution, our work enables optimizations that target not only total energy reduction but also efficiency at the level of kernels, phases, and full applications. This effort lays the foundation for runtime systems and analysis pipelines that can adapt to reliable energy measurements and support the next generation of fine-grained, power-aware optimization strategies.

\section{Frontier and Portage power and energy sensors}
\label{sec:frontier-portage-sensors}
\label{sec:platforms}

Modern HPE/AMD-based supercomputers expose a variety of on-chip and off-chip power and energy sensors that can be accessed by applications at runtime or via system telemetry. On Frontier (AMD Instinct\texttrademark~MI250X GPUs) and Portage (AMD Instinct\texttrademark~MI300A APUs), the sensors differ in both scope (for example, device versus node level) and behavior (for example, update interval or response time). A node-sensor layout for both systems is provided in the Appendix (Figure~\ref{fig:node-sensors}) for reference, but the methodology developed here depends on the sensor characterization described below rather than on a single hardware architecture.

\subsection{Sensor Operation and Interface}
\label{sec:monitoring_sensors}

\begin{figure}[tbp]
  \centering
  \includegraphics[width=\linewidth,height=0.55\textheight,keepaspectratio]{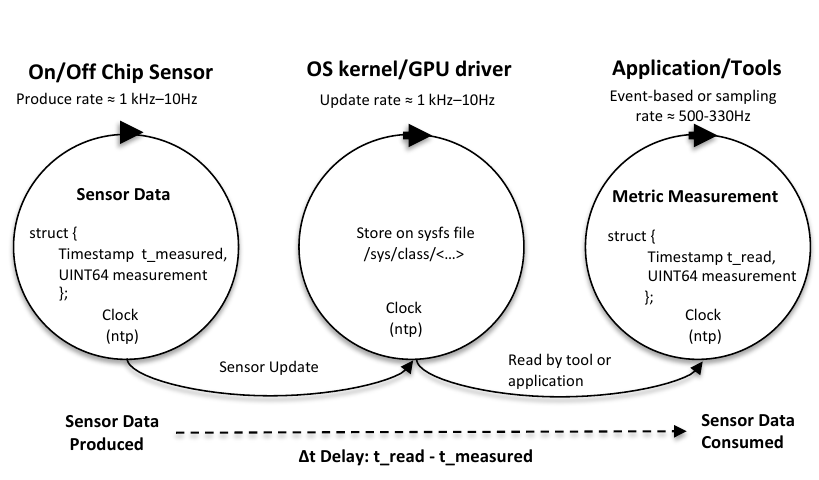}
  \caption{Three asynchronous stages for sensor production, driver or sysfs refresh publication, and application reads. A sensor update denotes publication of the most recent sensor value by the system layer. Application reads observe either that refreshed value or a cached value, but do not trigger a new measurement. The resulting lag is $\Delta t = t_{\text{read}} - t_{\text{measured}}$.}
  \label{fig:sensor-diagram}
\end{figure}

Figure~\ref{fig:sensor-diagram} illustrates the three asynchronous stages of the sensor pipeline. Each sensor produces measurements according to its own internal cadence and timestamp. The data then passes through two additional stages: (1) the system layer, where the OS and GPU driver refresh and publish values through the sysfs interface, and (2) the application layer, where tools such as PAPI and Score-P poll those published values at user-requested sampling rate.

We define \emph{sensor update} to denote the publication of the most recent sensor value by the system layer. Successive tool reads do not trigger new hardware measurements. They observe either a newly published value or the last cached value if no refresh has occurred since the previous read. Because sensor measurement, driver refresh, and application polling are asynchronous and rely on distinct timestamps, a delay ($\Delta t$) is introduced between the sensor reading and when it becomes visible to applications. NTP-based synchronization helps reduce drift, but small differences in timestamp resolution and update frequency remain at sub-millisecond timescales.

\subsection{On-chip Power and Energy with \texttt{rocm-smi}}
The ROCm System Management Interface (\texttt{rocm-smi} which is being replaced by \texttt{amd-smi}) is
the on-chip interface for GPU metrics. It provides power and energy sensors directly from the accelerator package through the GPU driver stack.
For this work, we employ the \texttt{rocm-smi} API to collect power and energy measurements for two types of devices on modern supercomputing platforms: the Instinct MI250X GPU (Frontier) and the Instinct MI300A APU (Portage).

The sensors on the Instinct MI250X GPU report power and energy values at the GPU package level.
Each physical GPU card on the MI250X contains two \emph{Graphics Compute Dies (GCDs)}, which are the functional dies responsible for executing kernels.
The \texttt{rocm-smi} reports aggregate power/energy values for the full GPU package, rather than for each GCD.
Thus, fine-grained differences between the two GCDs cannot be observed directly.
The sensors refresh cumulative energy values at 1\,ms granularity.
While the \texttt{rocm-smi} does not support instantaneous power measurements, average power based on a running average of energy values are supported, but the internal averaging window is not documented.

Each Instinct MI300A APU contains a set of Zen4 CPUs and CDNA3 GPUs that each share a unified HBM3 memory stack.
The \texttt{rocm-smi} aggregates power and energy measurements for the combined CPU, GPU, and memory activity on the APU and reports a single value for the full package rather than for each device, which can complicate attribution.
Similar to the MI250X, the MI300A supports energy and average power measurements at 1\,ms granularity.
Additionally, the MI300A supports a more fine-grained power counter, but its precise methodology, including any smoothing or signal processing, remains undocumented.

Although this study focused primarily on power and energy sensors, we provide the full \texttt{rocm-smi} sensor list we collected, including energy counts, clock frequencies, junction temperatures, and power, in Appendix Section~\ref{app:sensor-counters} (Tables~\ref{tab:rocm_frontier} and~\ref{tab:rocm_portage}).

\subsection{Node-Level Power and Energy with the Cray Power Management Counters}
\label{subsec:cray_pm}
HPE Cray EX systems expose power and energy telemetry at node-level and off-chip through Cray Power Management (PM) counters~\cite{martin2014cray,Martin2024craypm}.
These counters measure electrical input to major components on the node
motherboard before final point-of-load voltage conversions and therefore report
slightly higher values than on-chip sensors. Their accuracy has been validated
to within 5\% on Frontier (EX235a) and within 1\% on Portage (EX255a) based on testing during product development.

PM values are accessed through a Linux \texttt{sysfs} interface that refreshes
every 100\,ms. As described in Section~\ref{sec:monitoring_sensors}, this
refresh corresponds to the OS kernel-level update loop in the three-layer sensor
pipeline and introduces negligible overhead while providing stable node-level
telemetry. Additional technical details on PM counters, including their upstream measurement position, voltage-conversion, and step-down differences relative to on-chip sensors, hardware diagrams, and NIC-related offset behavior, are provided in Appendix~\ref{appendix:A}.

\paragraph{Frontier (EX235a).}
PM counters report node-level, CPU, memory, and per-GPU package power and energy for all four Instinct MI250X GPUs. Node-level totals include additional
components such as the Slingshot NICs, which draw from the baseboard and are
therefore folded entirely into the node counter rather than the per-GPU
counters. The complete list of PM counters available on Frontier is listed in the Appendix~\ref{appendix:A} Table~\ref{tab:pm_frontier}.

\paragraph{Portage (EX255a).}
Portage exposes PM counters at the node level and for each Instinct MI300A APU, whose
\texttt{accel[i]} counters report combined CPU+GPU+HBM package power. Two
of the four APUs share power rails with the Slingshot NICs, causing their PM
readings to include a small static offset (approximately 30\,W under idle
network conditions). This offset is removed during attribution. Because PM
sensors operate upstream of final voltage regulators, they systematically exceed
on-chip \texttt{rocm-smi} readings but remain consistent across iterations and suitable
for cross-sensor validation. The complete list of PM counters available on Portage is listed in Appendix~\ref{appendix:A} Table~\ref{tab:pm_portage}

\subsection{Tooling for Data Collection and Analysis}
\label{subsec:tooling}

We access these sensors and telemetry sources in-band from applications using  PAPI~\cite{jagode2024advancements} and the Score-P~\cite{10.1007/978-3-642-31476-6_7} performance measurement infrastructure.

\paragraph{PAPI and Score-P} 
We extended the \texttt{rocm-smi} component to expose 1,ms energy sensors on the MI250X and both energy and power sensors on the MI300A, in addition to the existing power sensor readings. We also integrated the Cray~PM sensors as additional PAPI 
components accessible directly from user space, providing unified in-band access 
to on-chip, off-chip, and node-level telemetry. Using the Async PAPI (APAPI) plugin we sample these sensors asynchronously in a 
dedicated thread per node per PAPI component, preventing interference with 
application threads. All sampled values are recorded in a unified tool time base and stored in the Open Trace Format Version 2 (OTF2), enabling the \textit{read} timestamp of each sensor sample to be used during post-processing to map metrics to code regions, processes, and devices at scale and to account for sensor measurement delays (Section~\ref{sec:monitoring_sensors}). The resulting synchronized traces are suitable for scalable statistical and time-series analysis because they preserve the relationship between application regions and heterogeneous sensor streams. The measured instrumentation overhead for collecting Score-P traces and polling PAPI counters is below 1\%, provided that additional cores are reserved for each Score-P plugin's sampling thread.

\paragraph{Trace Conversion and Preprocessing} 
A critical bottleneck in our workflow was trace analysis: converting OTF2 traces into CSV format for fine-grained, region-level power and energy attribution. Our initial Python-based workflow using the 
\texttt{otf2} module struggled with multi-gigabyte trace files, often taking 
longer to convert than to analyze. To overcome this limitation, we implemented a 
subset of the OTF2 library in Chapel~\cite{chamberlain2007parallel}, called 
\textit{fastotf2}~\cite{fastotf2}. This tool interfaces directly with the native 
OTF2 API and reconstructs call graphs and performance metrics for every thread 
and MPI rank in parallel, yielding an order-of-magnitude speedup in CSV 
generation compared to the Python-based module.

\paragraph{Power and Energy Trace Analysis} 
Following preprocessing, we load the generated CSV files into Pandas DataFrames within
Jupyter notebooks. By leveraging Python’s process-level parallelism we analyze hundreds
of traces concurrently using efficient row-wise operations across the call-graph and
metric DataFrames. This workflow enables rapid, scalable, and reproducible analysis of
large OTF2 datasets for sensor characterization and fine-grained power and energy
attribution for many nodes at once.

\section{Methodology}
\label{sec:methodology}

\subsection{Measurement Model}
\subsubsection{Sensor Characteristics}
\begin{figure}[htbp]
  \centering
  \includegraphics[width=\linewidth]{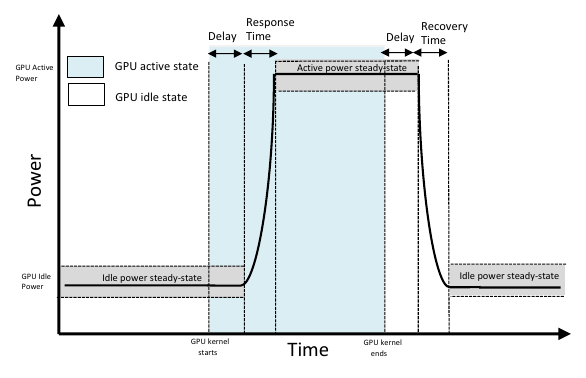}
  \caption{
GPU power sensor behavior during a kernel execution. The blue region shows the
active kernel interval; white regions denote idle periods. Vertical dashed lines
indicate delay, response, and recovery times; horizontal dashed lines show idle
and active steady states.
}
  \label{fig:gpu-response}
\end{figure}

\label{subsubsec:sensor_characteristics}

To evaluate the on-chip and off-chip power and energy sensors we characterize their behavior across several dimensions, including how each sensor reports values, how closely those values reflect hardware activity, and which characteristics can affect application-level attribution. The following subsections define these characteristics.

\paragraph{Update Interval.}
The update interval is the time between consecutive sensor readings and determines the temporal resolution of observable power changes. Short intervals enable fine-grained attribution, while long intervals can mask short-lived behavior. In our model (Section~\ref{sec:monitoring_sensors}), we distinguish between: (1) the sensor update interval, and (2) the interval at which a tool or application observes those updates. As shown in Figure~\ref{fig:sensor-diagram}, the sensor timestamp (\textit{t\_\text{measured}}) and the tool timestamp (\textit{t\_\text{read}}) are generated independently and are not synchronized. Differences in production cadence, driver refresh, sampling rate, latency, and caching determine the effective temporal resolution and must be characterized for accurate interpretation.

\paragraph{Delay, Response, and Recovery Time.}

Figure~\ref{fig:gpu-response} illustrates how GPU power sensors capture transitions between idle and active states. Three effects govern this behavior: \textit{delay}, \textit{response time}, and \textit{recovery time}~\cite{oppenheim2017signals}. \textit{Delay} is the latency between the true onset of a power change and the first observable update, caused by asynchronous interactions across device, driver, and sampling layers, which shift the measured trace and complicate fine-grained attribution. \textit{Response time} describes how quickly reported power rises after the delay; we use the 10--90\% rise-time metric~\cite{nise2019control}. Update intervals and filtering (e.g., averaging or low-pass smoothing) can elongate this rise and obscure short spikes. Prior work~\cite{yang2024accurate} reports patterns such as step changes, delayed transitions, linear ramps, and logarithmic behavior. \textit{Recovery time} is the 90--10\% decay to idle after kernel completion; the same filtering prolongs this decay, keeping reported power elevated beyond execution.

Together, these effects introduce temporal distortion: delay shifts event boundaries, while response and recovery smooth rising and falling edges.

For a phase with start and end times $t_s$ and $t_e$, we define a \emph{confidence window} for reliable attribution:
\begin{equation}
\mathcal{W}{\text{conf}} = \big[t_s + t_d + t_r,\;t_e - t_d - t_f\big]
\label{eq:confidence_window}
\end{equation}
where $t_d$ is delay, $t_r$ is the 10--90\% rise time (response), and $t_f$ is the 90--10\% fall time (recovery). Within $\mathcal{W}{\text{conf}}$, reported power approximates steady-state behavior; outside it, measurements are dominated by sensor transition effects.

\paragraph{Aliasing}

\begin{figure}[htbp]
    \centering
    \begin{subfigure}[b]{0.48\columnwidth}
        \centering
        \includegraphics[width=\linewidth]{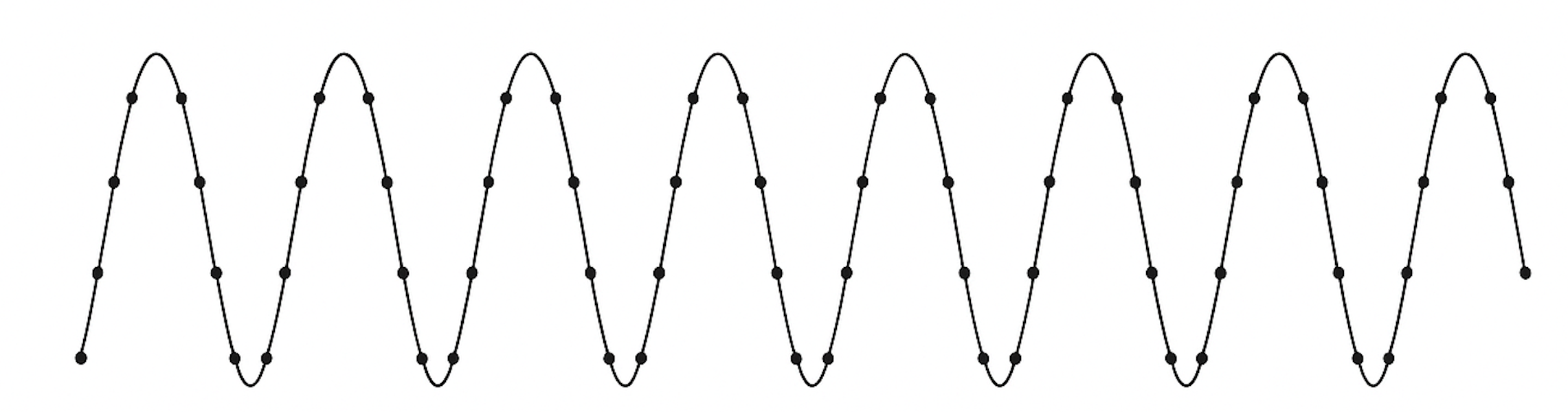}
        \caption{No aliasing}
        \label{fig:noalias}
    \end{subfigure}
    \hfill
    \begin{subfigure}[b]{0.48\columnwidth}
        \centering
        \includegraphics[width=\linewidth]{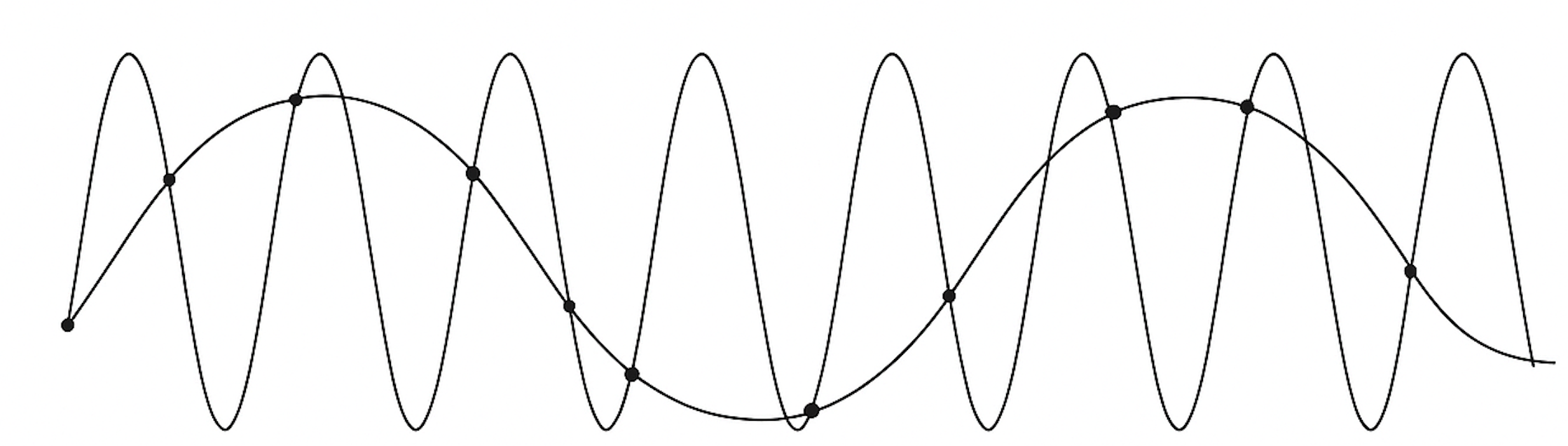}
        \caption{Aliasing}
        \label{fig:alias}
    \end{subfigure}
    \caption{Illustration of sampling effects: (a) signal captured without aliasing and (b) undersampled signal exhibiting aliasing distortion.}
    \label{fig:aliasing}
\end{figure}

Aliasing occurs when the update interval of a sensor is too coarse relative to the workload’s activity period. As a result, the reported values may be distorted, appearing smoothed or oscillatory in ways that do not reflect the true underlying signal. Figure~\ref{fig:aliasing} shows the effect of aliasing on an oscillatory signal. 

\paragraph{Thermal Design Power (TDP) and Power-Excursions Capture}
Power excursions refer to brief periods where power exceeds the nominal TDP during millisecond-scale bursts. A sensor’s ability to capture these events with good resolution, without clipping or excessive filtering and lag is critical for attributing transient overshoot to specific code regions and for evaluating hardware power-management responses.

\paragraph{Sensors corrections via offsets}
Offset and slope errors describe systematic deviations in the mapping between reported sensor values. In this particular case, we are applying an offset to correct the Cray PM Instinct MI300A APU power values. As shown in Fig. \ref{fig:node-sensors}, the packaging of MI300As on the motherboard is such that a Cassini Network card shares the power rails with two out of the four APUs, specifically APUs 0 and 2. These network cards draw a constant power of 30W based on our observations. Therefore, for data gathered from Portage, our measurements apply a 30W offset to the Cray PM counters corresponding to APUs 0 and 2.

\subsubsection{Deriving Instantaneous Power via $\Delta E / \Delta t$}
\label{subsec:deriving_power}
Modern GPUs report on-board power using moving-average filters that smooth short spikes and obscure fine-grained application behavior. On Instinct MI250X GPUs, for example, \texttt{rocm-smi} power values reflect a running average, and similar smoothing has been observed on other GPU architectures~\cite{yang2024accurate}. Such filtering can mask transient peaks over tens of milliseconds to one second and underestimate energy during short application events~\cite{yang2024accurate}. Although newer interfaces expose instantaneous power fields~\cite{burtscher2014,yang2024accurate}, several on-chip sensors still default to filtered values to provide smoother power readings.

\begin{figure*}[htbp]
  \centering
  \captionsetup[subfigure]{justification=centering,font=footnotesize,skip=0pt}

  \begin{subfigure}[htbp]{0.48\textwidth}
    \includegraphics[width=\linewidth,keepaspectratio,
                     trim=4 4 4 4,clip]{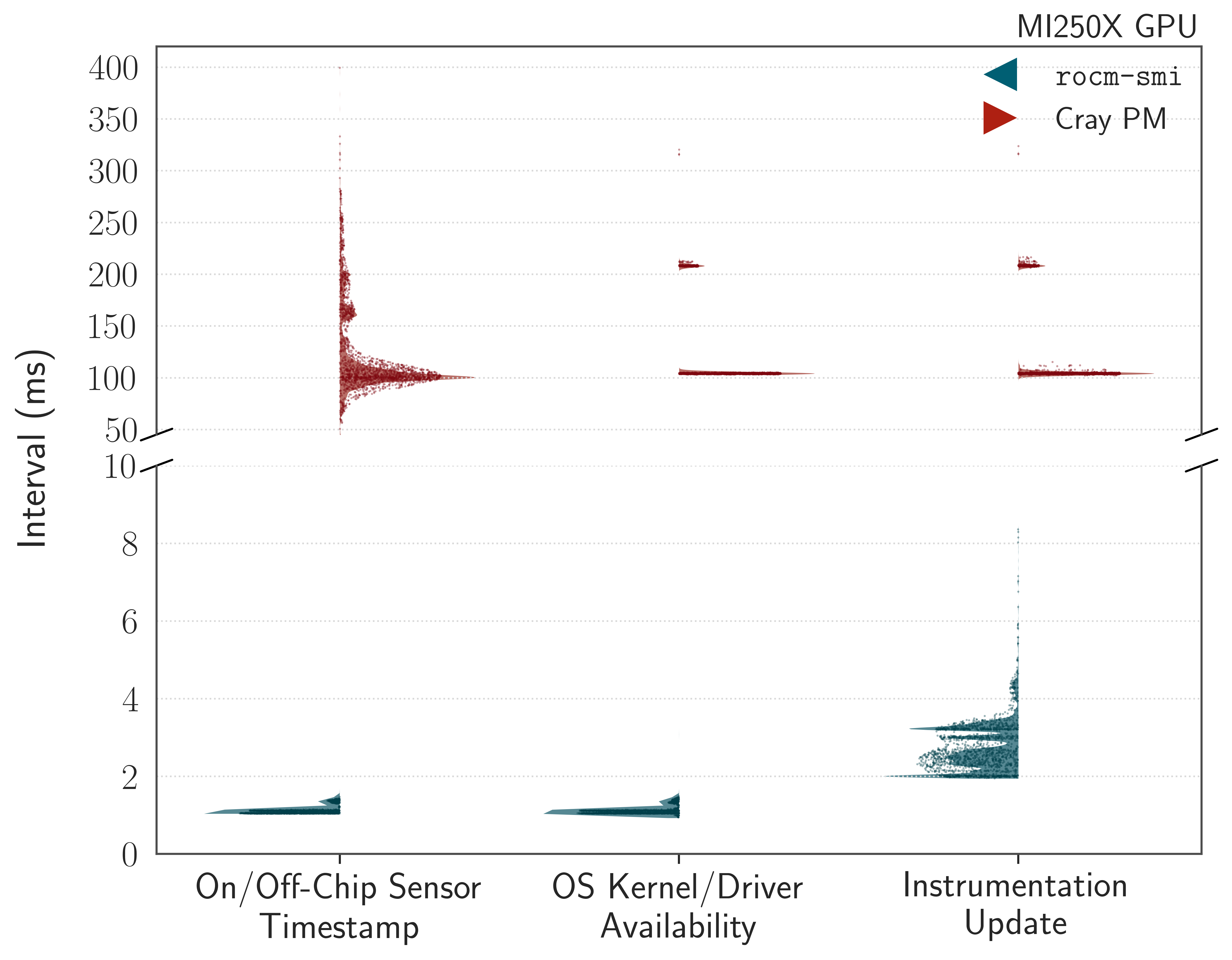}
    \caption{}
    \label{fig:update-interval-mi250x}
  \end{subfigure}\hfill
  \begin{subfigure}[htbp]{0.48\textwidth}
    \includegraphics[width=\linewidth,keepaspectratio,
                     trim=4 4 4 4,clip]{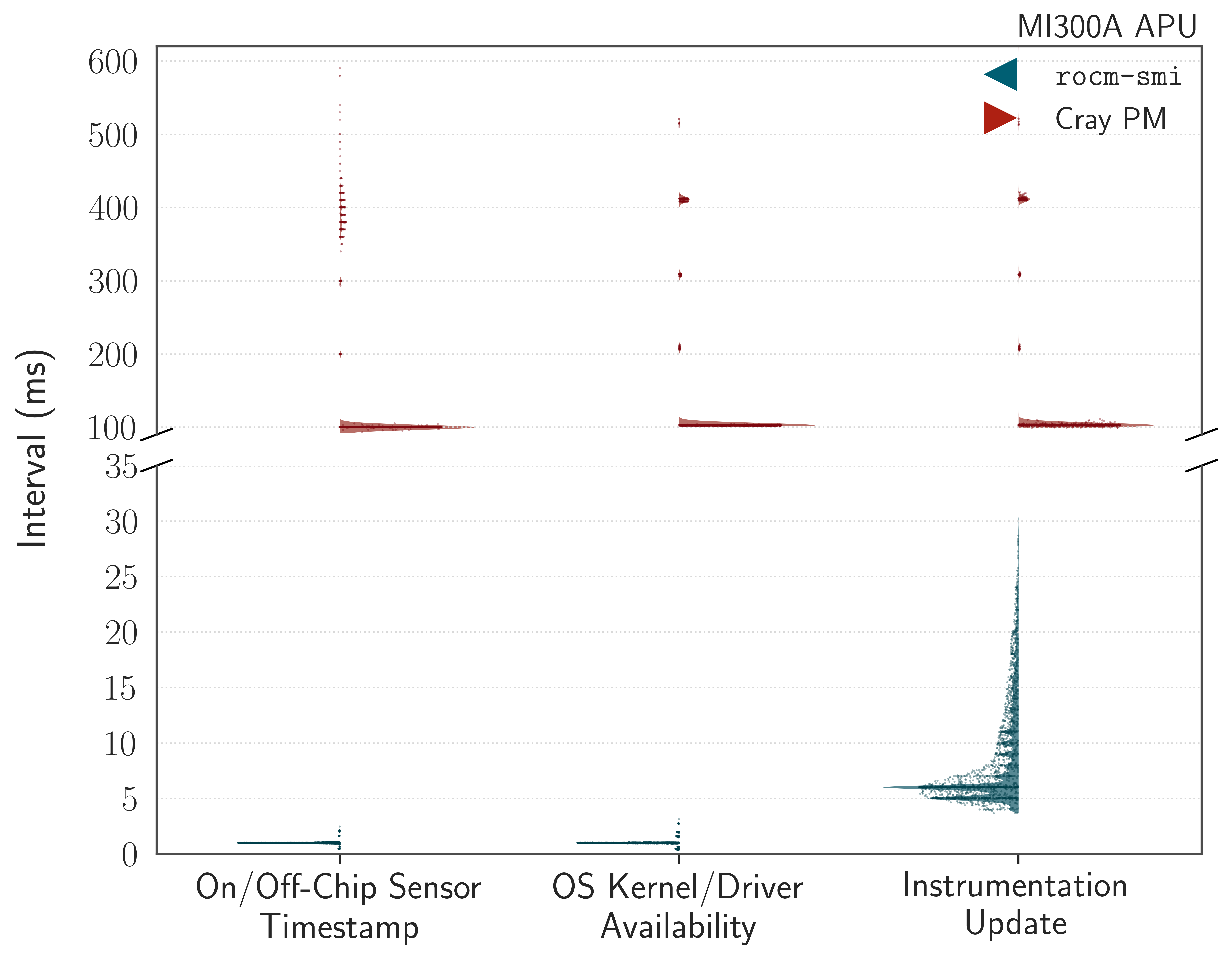}
    \caption{}
    \label{fig:update-interval-mi300a}
  \end{subfigure}

  \caption{Update-interval and timestamp-delta behavior measured across 128 nodes on Frontier (left column, 512 AMD Instinct MI250X GPUs) and Portage (right column, 512 AMD Instinct MI300A APUs). In each column, the top block shows consecutive $t_{\text{measured}}$ deltas reported by \texttt{rocm-smi} and Cray~PM, the middle block shows driver-visible sensor updates, and the bottom block shows value changes detected by the instrumentation layer. Together, these distributions separate sensor production cadence, driver publication cadence, and tool-observation cadence.}
  \label{fig:update-intervals-2x2}
\end{figure*}

To obtain higher temporal fidelity, we bypass these filters and derive instantaneous power directly from cumulative energy counters following prior methods~\cite{burtscher2014,yang2024accurate,jain2026minos}. AMD’s \texttt{rocm-smi}/\texttt{amd-smi}, NVIDIA \texttt{nvml}, and similar interfaces expose per-GPU energy counters that can be sampled at high frequency. We sample the cumulative energy counter for MI250X and MI300A at the lowest millisecond update intervals that our tool can measure (see Figure~\ref{fig:update-intervals-2x2}) and derive instantaneous power as
\[
\text{Power}_{\text{inst}}(i) \approx \frac{E(i)-E(i-1)}{\Delta t}.
\]

Energy-derived power reveals short transients that moving-average sensors smooth out. In our study, we evaluate this method by (1) comparing steady-state $\text{Power}_{\text{inst}}$ with averaged sensor values, and (2) examining rise and fall behavior during rapid transitions. Finally, cross-validation on MI300A APUs, which offer a native 1\,ms mode shows consistent trends across architectures and aligns with recent fine-grained profiling work such as FinGraV~\cite{singhania2024finegrain}.

\section{Experimental Setup}
\label{sec:setup}

\subsection{Target Systems}
The experiments were run on the Frontier supercomputer at the Oak Ridge Leadership Computing Facility (OLCF)~\cite{frontier-guide} and on Portage at Hewlett Packard Enterprise (HPE).

\subsubsection{Frontier}
Frontier is the exascale system at the Oak Ridge Leadership Computing Facility (OLCF), comprising 9,856 HPE Cray EX235a nodes. Each node includes one AMD EPYC 7A53 “Trento” CPU (64 Zen3 cores) and four AMD Instinct MI250X GPUs. Each MI250X is a dual-die device with two Graphics Compute Dies (GCDs), each acting as an independent accelerator with 64 GB of HBM2e (128 GB per GPU). GPUs are interconnected via AMD Infinity Fabric, which also links them to the CPU. Nodes are connected through HPE Slingshot (200 Gbps), with four Cassini NICs per node (two sawtooth cards). Frontier runs the HPE Cray Programming Environment, including Cray MPICH and ROCm 6.4.1 for GPU acceleration. Power and energy sensors are described in Section~\ref{sec:platforms}. Each GPU is bound to the peak power draw of 560 watts, the thermal design power limit, without a power cap. The dynamic frequency scaling (DVFS) policy is not changed from the GPU's default system. No benchmarks triggered thermal throttling, and for reference rocHPL showed an average GPU frequency of 1240 MHz while at TDP during the active region, which is throttled from the 1700 MHz peak frequency due to power constraints. The CPU uses DVFS with boost capability enabled.

\subsubsection{Portage}
Portage is an internal HPE cluster with 128 HPE Cray EX255a nodes, each containing four AMD Instinct MI300A APUs. Each APU integrates 24 Zen 4 CPU cores, 228 GPU compute units, and 128 GB of unified HBM3 memory. Unlike MI250X GPUs, which are split into GCDs, the MI300A exposes power and energy measurements at the APU level, combining CPU, GPU, and memory activity, which simplifies programming but complicates fine-grained energy attribution. APUs are interconnected via AMD Infinity Fabric, and nodes communicate through the HPE Slingshot 200 Gbps network. Each node includes four Cassini NICs (two sawtooth cards), with NIC power partially attributed to specific APU counters. The system uses the HPE Cray Programming Environment with ROCm 6.4.1. Each APU is power capped to 550 watts, with unchanged DVFS policy from the GPU's default. Similarly to Frontier, Portage showed no thermal throttling, and rocHPL exhibited an average GPU frequency of 1015MHz while at TDP, throttled from the 2100 MHz peak frequency.

\subsection{Synthetic Square-Wave Workloads}
\label{subsec:synthetic}
To characterize the power-monitoring methodology presented in this work, we developed a GPU-based workload that alternates between two states: one in which the GPU is fully active, drawing power at the TDP level, and another in which the GPU is inactive, drawing power at the idle state level. The durations of the active and inactive states are user-specified, as is the number of iterations for the alternating sequence. During the active regions, the workload launches multiple instances, one after the other, of a kernel that performs several double-precision vector fused multiply-add (FMA) operations. The number of FMAs performed by the kernel has been calibrated such that the data movement rate from HBM is close to the computation rate, therefore heavily exercising the GPU's bandwidth and compute resources and driving the power usage to the TDP limit. We refer to this benchmark as the \textit{square-wave workload}. When running on multiple GPUs, the benchmark leverages MPI to synchronize the GPUs so that all GPUs start each active interval at the same time.  

\subsection{Benchmarks: HPL and HPG-MxP (Full and Mixed Precision)}
\subsubsection{rocHPL}
rocHPL~\cite{10.1145/3581784.3607066} is AMD’s ROCm-based native implementation of the LINPACK benchmark. It solves a dense linear system (Ax=b) using LU factorization with partial pivoting and reports sustained FP64 performance (GFLOP/s to TFLOP/s) as a function of problem size and process grid. It is built on ROCm libraries (rocBLAS, rocSOLVER, RCCL for multi-GPU communication) and exercises the classic “HPL phases” like panel factorization, updates, and communications. It is a good baseline for studying power/energy during FP64-intensive compute plus collectives.

\subsubsection{rocHPL-MxP}
rocHPL-MxP~\cite{rochpl_mxp_github} is the mixed-precision variant of HPL, factoring in low precision (e.g., FP16/BF16 with FP32 accumulations via tensor/matrix cores where available) and uses iterative refinement to recover FP64-accurate solutions. Apart from the usage of mixed precision dense matrix operations, another significant difference from HPL is that no pivoting is performed nor required since the matrix is specially constructed to be diagonal-dominant. In our study, these mixed-precision phases help expose attribution pitfalls and the benefit of instantaneous reconstruction.

\subsubsection{HPG-MxP}
HPG-MxP stresses the memory hierarchy with sparse linear-algebra phases across precisions \cite{hpgmp,kashi2025scaling}. It is useful when one wants: (a) stable, repeatable memory bandwidth-bound segments for sensor analysis, (b) knobs for problem size/precision that change power signatures, and (c) clear phase boundaries for attribution (plan/allocate, warm-up, steady iterative solver loops, finalize). In our results we report instantaneous power per phase and show how short phases can be under- or over-attributed if you don’t compensate for sensor lag/averaging.
It may be noted that while HPL and HPL-MxP are two separate benchmarks, HPG-MxP is a single benchmark that contains both double-precision and mixed-precision runs for comparison.

\section{Results}
\label{sec:results}

\subsection{Sensor Characterization via Square-Waves}

\subsubsection{Sensor Update Intervals}

As discussed in Section~\ref{sec:monitoring_sensors}, two update intervals
matter for attribution: the native update interval and the interval
at which the instrumentation layer detects those updates. Figure~\ref{fig:update-intervals-2x2} makes this distinction explicit: the left panel corresponds to Frontier and the right panel to Portage; within each panel, the left column shows consecutive \textit{t\textsubscript{measured}} deltas reported by the sensors, the center column shows sensor updates made visible through the kernel or driver layer, and the right column shows value changes actually observed by the instrumentation layer. The top blocks (shown in red) in Figures~\ref{fig:update-interval-mi250x} and~\ref{fig:update-interval-mi300a} show that Cray PM updates are more variable and occasionally stretch beyond the nominal cadence, whereas \texttt{rocm-smi} (shown in blue) updates are more tightly clustered. This maximum achievable refresh rate determines how predictably new power measurements can be made available to the tool for attribution.

\begin{figure*}[htbp]
  \centering
  \begin{subfigure}[t]{0.49\textwidth}
    \centering
    \includegraphics[width=\linewidth,height=0.55\textheight,keepaspectratio]{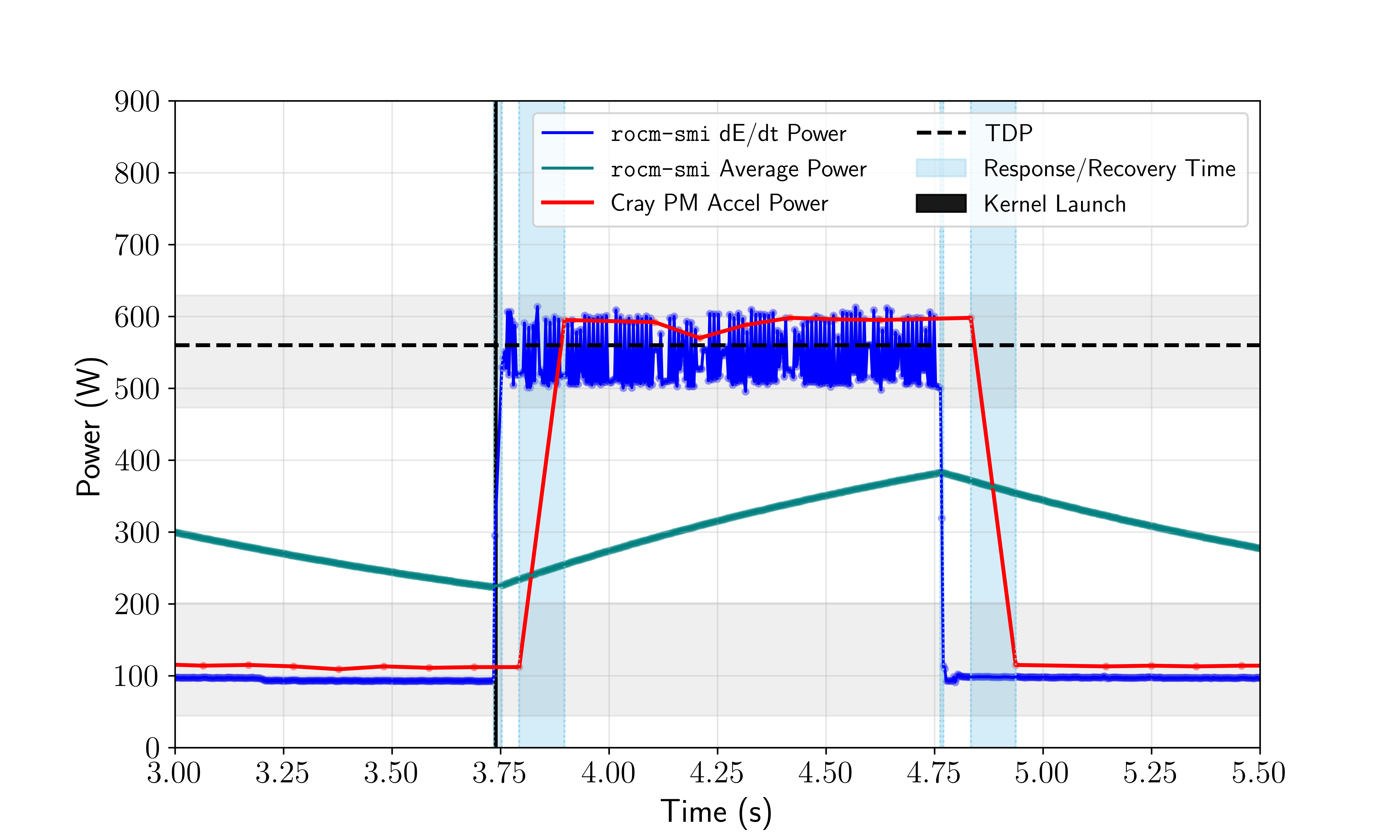}
    \caption{Power measured on an AMD Instinct\texttrademark~MI250X for a 1~s idle / 1~s active square-wave workload. Averaged \texttt{rocm-smi} power, derived $\Delta E / \Delta t$ power, TDP, and \texttt{cray\_pm} illustrate delay, response, and recovery behavior with relatively sharp transitions.}
    \label{fig:mi250x-response-time}
  \end{subfigure}
  \hfill
  \begin{subfigure}[t]{0.49\textwidth}
    \centering
    \includegraphics[width=\linewidth,height=0.55\textheight,keepaspectratio]{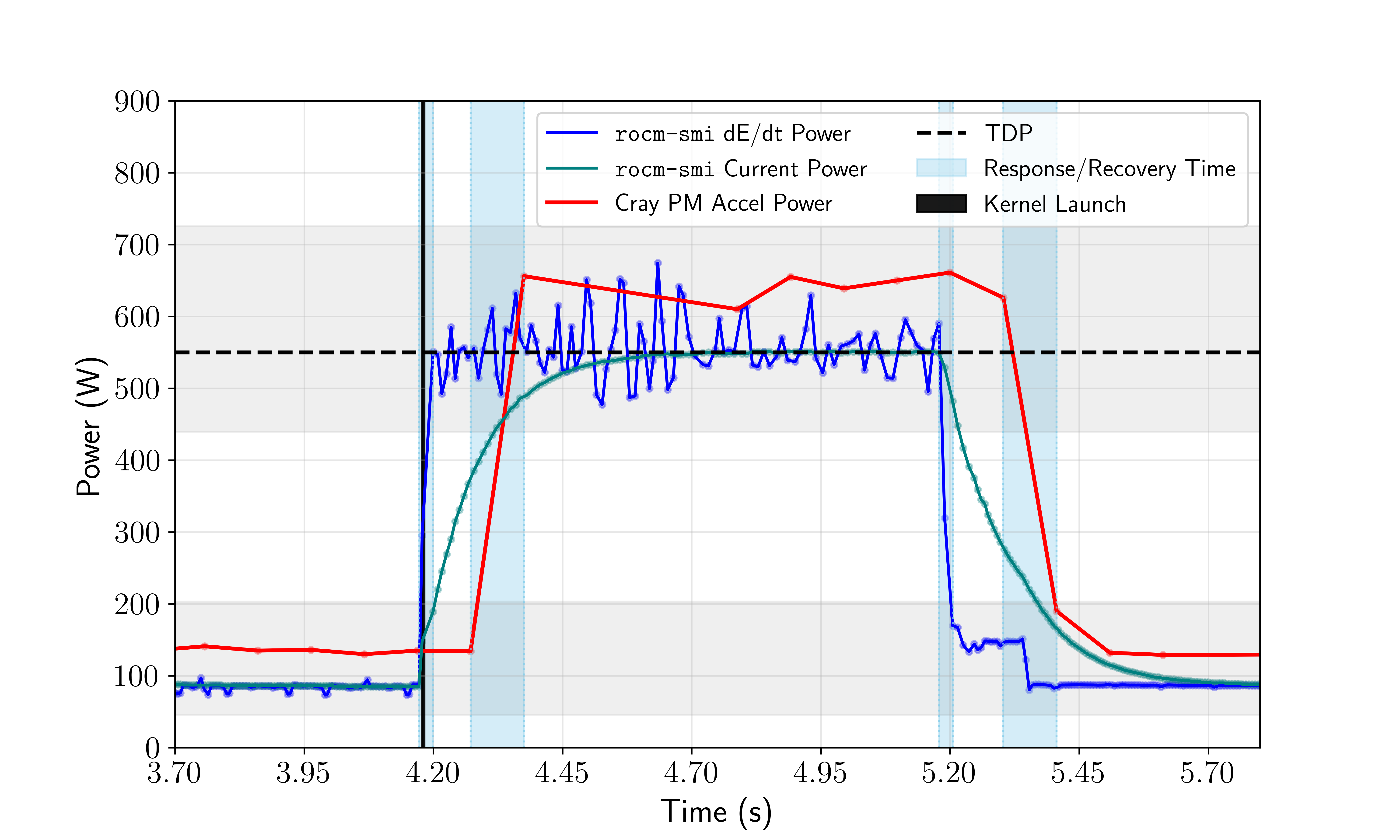}
    \caption{Power measured on an AMD Instinct\texttrademark~MI300A under the same 1~s idle / 1~s active square-wave workload. Averaged current power exhibits stronger smoothing on rise and fall, while derived $\Delta E / \Delta t$, TDP, and \texttt{cray\_pm} confirm consistent levels across sensors.}
    \label{fig:mi300a-response-time}
  \end{subfigure}
  \caption{Delay, response, and recovery behavior for AMD Instinct\texttrademark~MI250X (left) and AMD Instinct\texttrademark~MI300A (right) under a square-wave GPU load, comparing filtered ROCm-SMI measurements, derived $\Delta E / \Delta t$ power, and off-chip \texttt{cray\_pm}.}
\end{figure*}

The rightmost column in Figures~\ref{fig:update-interval-mi250x} and~\ref{fig:update-interval-mi300a} show the distribution of \textit{t\textsubscript{read}}, which reflects what the instrumentation actually observed during sampling. These intervals incorporate both native sensor timing and the overhead associated with capturing and logging values in the tracing tool (Score-P via PAPI). In our experiments, this overhead comes primarily from sampling 24 sensors across all four GPUs in a node (see Appendix Tables~\ref{tab:rocm_frontier} and~\ref{tab:rocm_portage}), which introduces jitter into \textit{t\textsubscript{read}}, especially for \texttt{rocm-smi}, where the tool attempts to sample every 1~ms. This additional variability appears as a wider distribution and occasional longer gaps between observed updates. For Cray PM, the instrumentation layer generally captures all available refreshes, although individual updates may still be missed if two native updates occur between consecutive reads.

These results show that the interval produced by the sensors (\textit{t\textsubscript{measured}}) and the interval observed by the instrumentation layer (\textit{t\textsubscript{read}}) can differ significantly. The native interval measures how often new sensor data is produced, while the observed interval reflects how often those updates are captured by the tool. For Cray PM, the dominant bottleneck is the 100\,ms driver refresh cadence, so polling faster does not reveal faster hardware updates; the application may simply read the same cached value multiple times. For \texttt{rocm-smi}, the situation is reversed: the on-chip energy counter refreshes at millisecond scale, but the instrumentation layer becomes the bottleneck because of the cost of sampling and logging many metrics per GPU. This distinction between sensor refresh limits and tool-observation limits is central to the aliasing behavior discussed later in Section~\ref{sec:results}.

The left columns in each panel of Figure~\ref{fig:update-intervals-2x2} also expose variability in the sensor-associated timestamp itself. Unlike the center and right columns, which describe refresh publication and tool detection, the left columns describe when the sensor reports that a measurement was taken. Even when the reported values change regularly, these measurement timestamps are not perfectly uniform. This variability must be taken into account when aligning sensor measurements with the application timeline because both the values and their associated timestamps fluctuate.

\begin{figure}[htbp]
  \centering
\includegraphics[width=\linewidth,height=0.7\textheight,keepaspectratio]{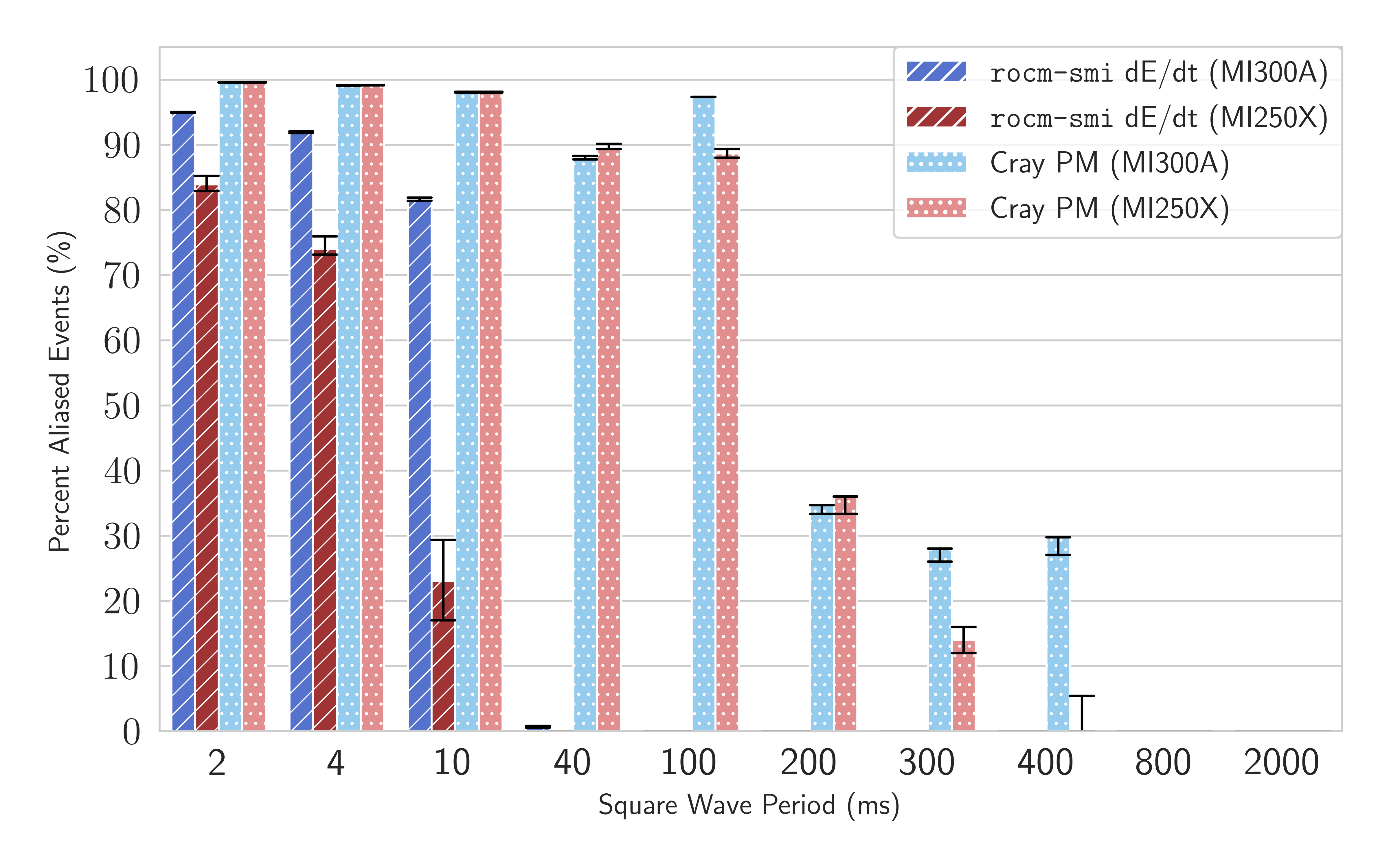}
  \caption{Median impact of aliasing on power state transition detection, shown with 95\% confidence intervals calculated across 512 MI250X GPUs and 480 MI300A APUs.}%
  \label{fig:frontier-portage-aliasing}
\end{figure}
\begin{figure*}[htbp]
  \centering
  \captionsetup[subfigure]{justification=centering}

  \begin{subfigure}[t]{0.495\textwidth}
    \centering
    \includegraphics[width=\linewidth,keepaspectratio,
                     height=0.49\textheight]{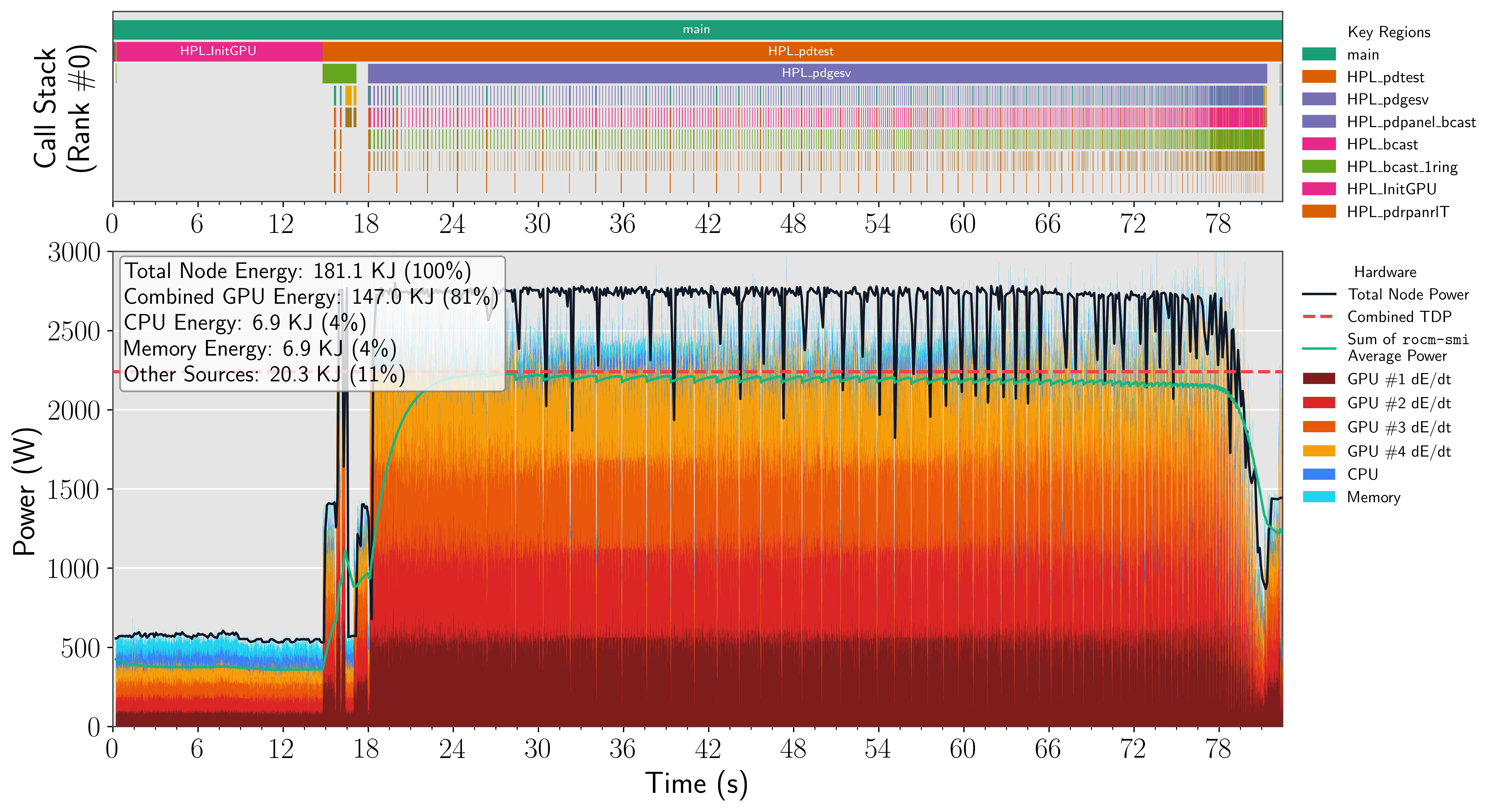}
    \caption{Frontier rocHPL (Full Precision): stacked instantaneous power per node component over time. GPU package power dominates during compute phases, while CPU, memory, and NIC contribute a smaller, mostly static baseline.} %
    \label{fig:frontier-hpl-stacked-power}
  \end{subfigure}
  \hfill
  \begin{subfigure}[t]{0.495\textwidth}
    \centering
    \includegraphics[width=\linewidth,keepaspectratio,
                     height=0.49\textheight]{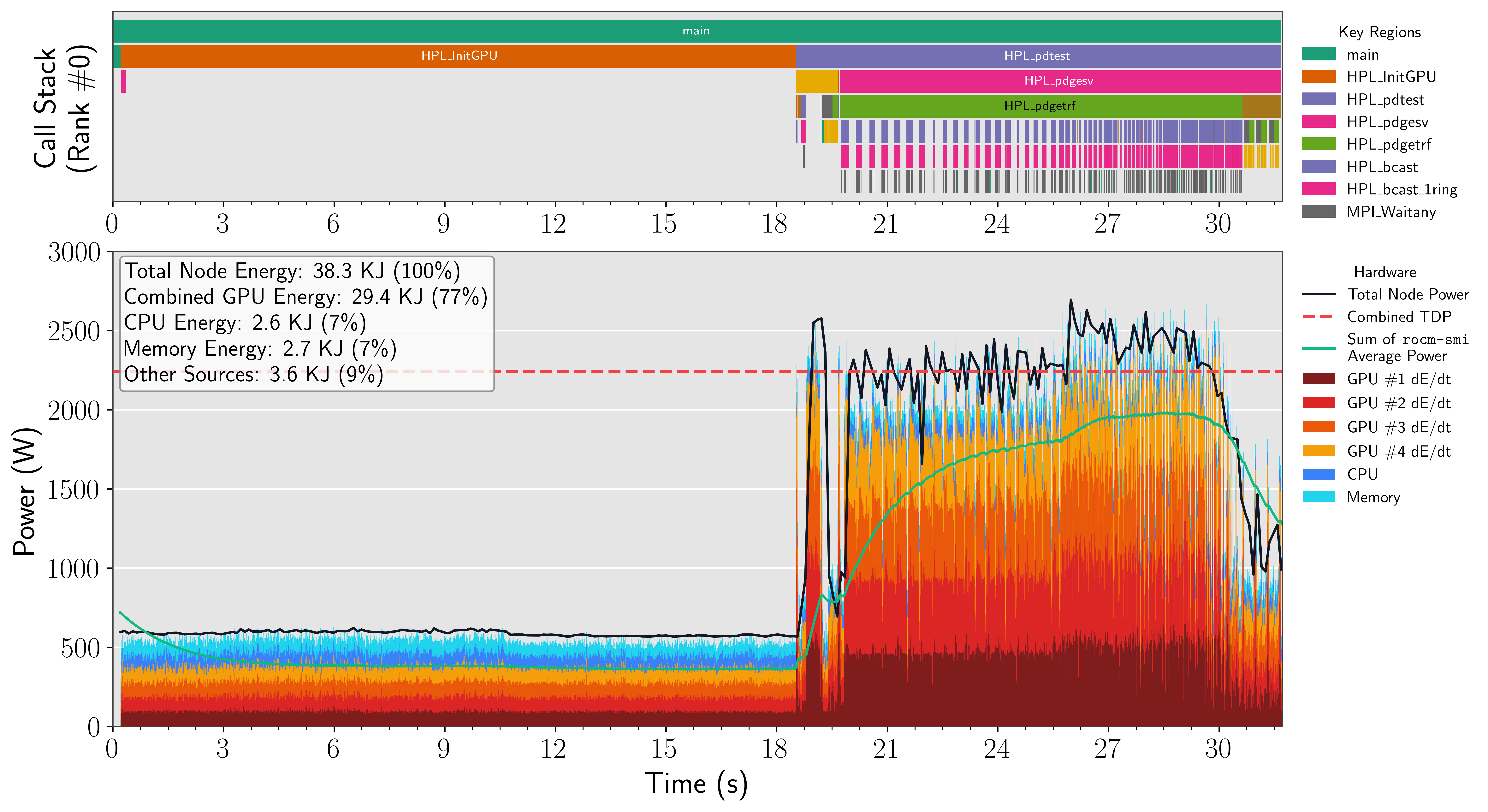}
    \caption{Frontier rocHPL-MxP (Mixed Precision): stacked instantaneous power per component. Power distribution mirrors the full-precision case, but the execution window is significantly shorter, reflecting mixed precision’s speedup.} %
    \label{fig:frontier-hplmxp-stacked-power}
  \end{subfigure}

  \caption{Frontier (AMD Instinct MI250x) rocHPL and rocHPL-MxP: side-by-side comparison of stacked instantaneous power.
  }
  \label{fig:frontier-hpl-stacked-combined}
\end{figure*}

\subsubsection{Delay, Response, and Recovery Time}

To quantify delay, response, and recovery behavior on production hardware, we
drive each device with a square-wave workload that alternates between 1~s
idle and 1~s active phases on the GPU. During the idle phase, no kernels are
issued; during the active phase, a compute-intensive kernel saturates the
device. For the Instinct MI250X experiments, we launch work on both GCDs and record
the \emph{averaged} package power reported by the vendor interface,
aggregated across the two GCDs. For the Instinct MI300A APU, we record the
\emph{averaged} current power reported by the APU interface. In both cases,
the plotted signals reflect filtered, moving-average views of device power
rather than instantaneous measurements. These traces allow us to observe the
three timing components introduced earlier: delay (initial lag), response
time (10\%--90\% rise), and recovery time (90\%--10\% fall).

Figure~\ref{fig:mi250x-response-time} shows the resulting trace for an
Instinct MI250X GPU under this 1~s idle / 1~se active square wave. Although the
workload transitions are effectively step changes at the hardware level, the average power reported by \texttt{rocm-smi} (green line) does not capture the full transition from idle to TDP, even in the one second step-wave. This is because the average power reported by \texttt{rocm-smi} is filtered at the firmware level. In this case, due to averaging, it takes a few seconds to fully capture the transition from idle to TDP, which is not useful for millisecond-level region attribution.
   
After the kernel phase begins, the power sensors remain at
the idle level for a short delay interval before rising toward the active
plateau. This rising portion spans multiple sensor update intervals, directly reflecting the combined effect of device-level power integration and the averaging applied in the monitoring stack, and thus defining the \emph{response time}. When the active phase ends the averaged package power
remains elevated for a comparable interval before decaying back to the idle
steady state. This trailing portion defines the \emph{recovery time}. Because
each phase is only 1~s long, delay, response time, and recovery time consume
a non-trivial fraction of the phase duration, reducing the portion of each
phase during which the averaged power is fully settled.

Figure~\ref{fig:mi300a-response-time} shows the analogous 1~s idle / 1~s
active experiment on an Instinct MI300A APU. The overall behavior is similar: the
measured power lags behind the true workload transitions (delay), then rises
gradually toward the active steady state (response), and finally decays back
toward idle with a visible tail once the work completes (recovery). However, on MI300-based systems the current power reported by \texttt{rocm-smi} is filtered over a smaller time window compared to the average power reported on MI250X. As shown in Figure~\ref{fig:mi300a-response-time}, the current power (green line) takes only $\sim$0.5~s to capture the idle to TDP transition, making it a closer measurement of the instantaneous power but still smoothing high-frequency fluctuations. Because of this smoothing, for square waves with active/idle times $\lesssim$0.5~s the average power
does not reach a perfectly flat plateau within the active window and the
decay into the subsequent idle phase overlaps with the next rising edge when
the square wave repeats. At the sensor level, neighboring phases are thus
partially merged, even though the underlying kernel schedule remains strictly
periodic.

Both figures also overlay our \emph{derived} GPU power, computed as
$\Delta E / \Delta t$ from the high-resolution energy counters. In contrast
to the \texttt{rocm-smi} averaged and current power signals, this derived power
shows shorter delay and response times with sharper transitions that more
closely track the start and end of the GPU-active intervals, and a faster
return to the idle level after the work completes. This makes the derived
power particularly well suited for attributing power and energy to kernels
and short phases, since the onset and termination of activity are preserved
with minimal temporal smearing.

To provide additional context, we include the GPU thermal design power (TDP)
and the off-chip power reported by \texttt{cray\_pm} in both plots. As
expected for an off-chip sensor, the \texttt{cray\_pm} readings
are consistently higher than the device TDP, reflecting contributions from
other components and conversion losses. At the same time, the overall
agreement between the filtered \texttt{rocm-smi} signals, our derived power, and the
off-chip \texttt{cray\_pm} measurements indicates that the $\Delta E /
\Delta t$ derivation appears well calibrated: the derived power tracks GPU
activity with better temporal resolution, while remaining consistent with the
absolute levels indicated by both the on-device averaged power and the
off-chip sensor.

When compared with the Cray PM power readings, the derived power from the \texttt{rocm-smi} energy counters is consistently lower than the Cray PM measurements by about 9\% on Frontier across all benchmarks. This gap is consistent with the expected 5\%-10\% losses due to the difference in placement of the sensors: the off-chip Cray PM readings are taken outside of the GPU package before a large voltage step-down, while \texttt{rocm-smi} reports on-chip measurements taken inside the package.

\subsubsection{Aliasing}

  \renewcommand{\arraystretch}{0.9}

To investigate the impact of aliasing on metric fidelity, we study the response of each sensor to the square-wave signal outlined in Section~\ref{subsec:synthetic}. We limit the analysis to equal idle and active intervals, so a square wave with a period of 2\,ms consists of 1\,ms of active time and 1\,ms of idle time. The workload was executed on 128 Frontier nodes (with 4 instances per node, each dispatching kernels to 2 GCDs on the same MI250X GPU) and on 120 Portage nodes (with 4 instances per node, each running on a separate APU). We collected off-chip instantaneous power readings from Cray PM sensors and calculated the on-chip instantaneous power with \texttt{rocm-smi} using successive energy readings as outlined in Section~\ref{subsec:deriving_power}. Figure~\ref{fig:frontier-portage-aliasing} reports the median power-state transition detection error across all nodes on each system for different square-wave periods, together with the 95\% confidence interval at each value. A sensor is considered to have recorded an active state when the measurement exceeds the average power for that node during the square-wave run, and an idle state otherwise.
Figure~\ref{fig:frontier-portage-aliasing} shows that the aliasing cutoff is governed by the slowest stage needed to resolve a transition. The sensor update interval determines the aliasing cutoff for the physical sampling frequency: the on-chip sensors measure at 1ms intervals, so the theoretical upper limit for sampling events without aliasing is 500Hz, the Nyquist frequency. If the workload's power usage changes at a higher frequency, the fluctuations in power draw are not captured in the measurements passed to the driver.

The instrumentation update interval creates another layer of aliasing: if the instrumentation consumes the sensor's data slower than it is produced, the instrumentation introduces downsampling. This second layer of aliasing could skip peaks and troughs in the power usage that the hardware might have originally measured.
The moving average in \texttt{rocm-smi} natively-reported power metric introduces a low pass filter which diminishes the high frequency components of the raw signal (i.e., it blurs the effect of short lived, burst like events). This filtering shifts the aliasing cutoff to longer intervals making the sensor appear slower to respond. For Cray PM, the dominant limit is the native driver refresh cadence, so transitions shorter than roughly 100\,ms are frequently missed even with faster polling. For \texttt{rocm-smi}, the native energy counter refreshes at millisecond scale, but the effective detection interval is widened by the costs of sampling and logging, which is why aliasing begins below roughly 4\,ms on MI250X and at longer periods on MI300A. Firmware smoothing on vendor-reported power fields further suppresses short transitions, which is why we base this analysis on $\Delta E / \Delta t$ rather than the averaged power counter. These observations distinguish sensor refresh limits from tool-side observation limits and explain why similar workloads exhibit different aliasing thresholds across the two platforms.
We also observe that aliasing for \texttt{rocm-smi} extends further on MI300A than on MI250X. This effect is consistent with the longer effective detection interval on MI300A, which reduces the rate at which the instrumentation layer can observe distinct state transitions.

FFT analysis can also be used to identify aliasing behavior. In the absence of aliasing, the square wave workload's power signal shows clear, distinct peaks at the square wave's harmonic frequencies. When aliasing is present, components above the Nyquist frequency are undersampled and folded back into lower frequencies. These components appear as broadband spectral noise, creating artifacts outside of the square wave's true harmonics and shifting the FFT's anticipated peaks. A visualization demonstrating this phenomenon is provided in the Appendix, in Figure~\ref{fig:fft-spectrum}.

\subsection{Benchmark Studies}
\label{sec:apps}
\subsubsection{Frontier: rocHPL, rocHPL-MxP and HPG-MxP}
\label{sec:bench-rochpl-frontier}

In this section we apply our sensor characterization and attribution
methodology to production workloads: the ROCm-based implementations of
the High Performance Linpack benchmark (rocHPL), its mixed-precision
variant (rocHPL-MxP), and the HPG-MxP sparse linear algebra benchmark in both
full-precision and mixed-precision modes. We run these benchmarks on
Frontier (Instinct MI250X GPUs) and Portage (Instinct MI300A APUs) and use the
multi-sensor tracing infrastructure described in Sections~\ref{sec:frontier-portage-sensors}
and~\ref{sec:methodology} to obtain millisecond-scale power and energy
profiles.
These case studies are intended to demonstrate the methodology under realistic application behavior rather than to claim novelty in the benchmarks themselves. Our goals are to (i) verify that $\Delta E / \Delta t$-derived power remains consistent with independent Cray PM measurements within steady-state confidence windows, (ii) show that phase-level attribution separates reduced time-to-solution from changes in instantaneous power, (iii) demonstrate that the same node-centric workflow applies to both discrete GPUs and integrated APUs, and (iv) generate synchronized traces that are suitable for downstream statistical and time-series analysis.

\begin{figure*}[htbp]
  \centering
  \captionsetup[subfigure]{justification=centering}
  \begin{subfigure}[t]{0.495\textwidth}
    \centering
    \includegraphics[width=\linewidth,height=0.49\textheight,keepaspectratio]{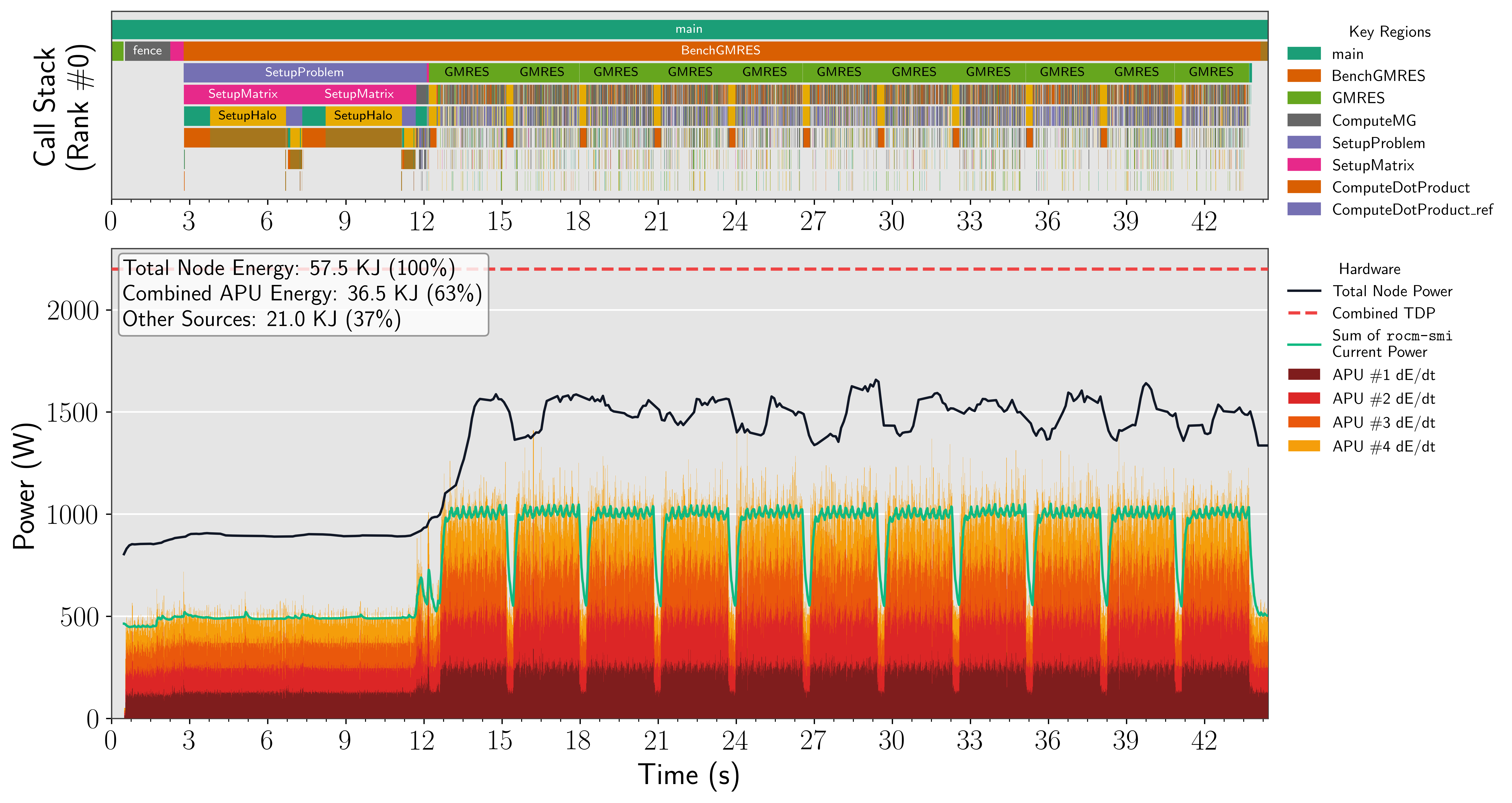}
    \caption{Portage HPG-MxP (Full Precision): stacked instantaneous power per component. 
    Compute and Krylov phases yield accelerator-heavy power signatures, 
    while communication segments slightly reduce APU power.}
    \label{fig:portage-hpg-stacked-power}
  \end{subfigure}\hfill
  \begin{subfigure}[t]{0.495\textwidth}
    \centering
    \includegraphics[width=\linewidth,height=0.49\textheight,keepaspectratio]{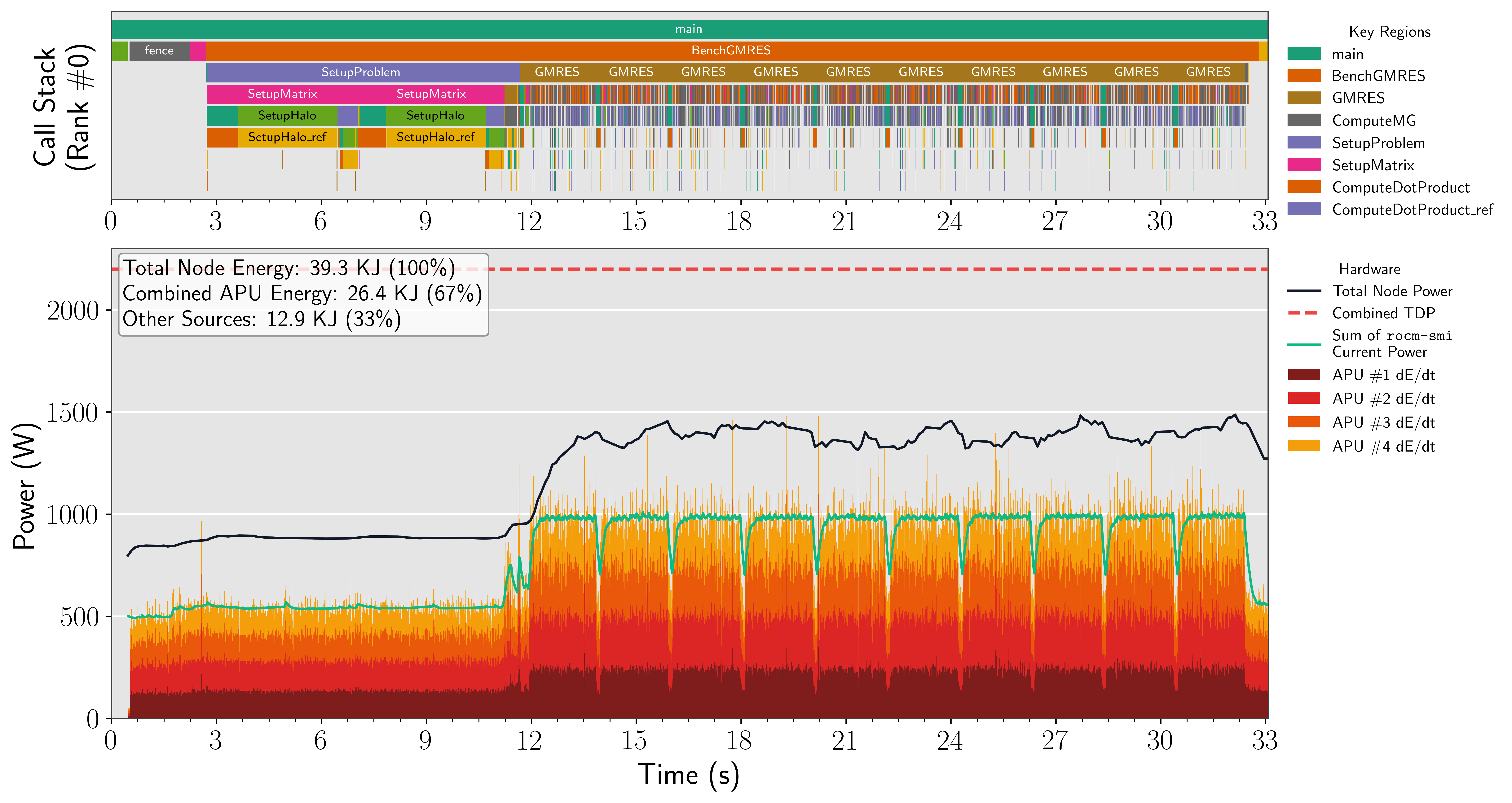}
    \caption{Portage HPG-MxP (Mixed Precision): stacked instantaneous power per component. 
    Mixed precision reduces runtime and node energy while preserving the qualitative 
    power distribution across APUs and node components.}
    \label{fig:portage-hpgmxp-stacked-power}
  \end{subfigure}

  \caption{Portage (AMD Instinct MI300A) HPG-MxP full-precision and mixed-precision stacked instantaneous power 
  traces shown side-by-side for comparison. 
  }
  \label{fig:portage-hpg-1x2}
\end{figure*}
\subsubsection{Benchmark Configuration and Instrumentation}
\label{sec:bench-setup}

For each benchmark we run a node-level configuration that exercises all
four accelerators available on the node: four Instinct MI250X GPUs on Frontier
and four Instinct MI300A APUs on Portage. We use the vendor-provided ROCm-based
implementations of rocHPL and rocHPL-MxP for HPL, and HPG-MxP in both
full-precision and mixed-precision modes \cite{kashi2025scaling}. Problem sizes and
process grids are chosen such that the full-precision variants saturate
device memory, while the mixed-precision runs use the same problem
sizes to allow a direct comparison of time-to-solution and energy.

Benchmark instrumentation is performed using Score-P with PAPI plugins that read on-chip, off-chip, and node-level sensors, and with the Score-P HIP plugin to instrument program regions, HIP events, and GPU kernels. Regions corresponding to major benchmark phases: initialization, matrix setup and distribution, panel factorization, trailing updates, iterative refinement, Krylov iterations, MPI collectives, and finalization, are annotated in the application, producing an OTF2 trace with a unified timestamp space for all CPU and GPU activities. In parallel, the two PAPI plugins sample \texttt{rocm-smi} and Cray PM sensors in dedicated threads on each node, avoiding interference with application threads.

On AMD GPUs and APUs we use Score-P to collect the cumulative energy counter every few milliseconds and derive instantaneous power as $\Delta E / \Delta t$ as described in
Section~\ref{sec:methodology}. This removes undocumented moving-average
filters from the vendor ``power'' counters and yields response and recovery times of a few milliseconds for the transitions
at phase boundaries. For node- and device-level off-chip measurements we use the Cray~PM counters at 10 Hz. On Portage, we apply the
30\,W offset for APU~0 and APU~2 Cray~PM readings to remove the static
Cassini NIC contribution, as discussed in Section~\ref{sec:frontier-portage-sensors}. 

All sensor streams are aligned offline using their timestamps and then
integrated over the Score-P phase intervals. This produces a consistent
set of phase-level power and energy metrics for each accelerator, each
node component (CPU, memory, NIC), and the full node.
Because sampling and region annotations are recorded independently on each node, the same methodology extends naturally to multi-node runs: Score-P records MPI regions across ranks, each node contributes aligned sensor streams, and system-level power or energy is obtained by summing node-level traces over common intervals. We do not add new multi-node results here, but the methodology is not limited to single-node analysis.

Figure~\ref{fig:frontier-hpl-stacked-combined} compares the full-precision rocHPL trace on the left with the mixed-precision rocHPL-MxP trace on the right. The stacked node-component power traces resolve a short initialization and matrix-distribution stage, a long compute-dominated plateau, and a brief tail, showing that the attribution is phase aware rather than only whole-program. In both cases, the four Instinct MI250X GPUs dominate node power during the main compute region, while CPU, memory, and NIC power form a relatively flat baseline.

The mixed-precision trace draws comparable instantaneous GPU power during the main compute region, but that region is much shorter. This allows the methodology to separate two effects that are often conflated: most of the energy reduction comes from reduced time-to-solution, while the change in instantaneous power is comparatively small. Across 128 independent runs on 128 nodes, the average node energy consumption was 179.9\,KJ for rocHPL and 38.5\,KJ for rocHPL-MxP, with standard deviations of 2.2\,KJ and 0.7\,KJ, respectively, corresponding to an average reduction of 79\%.

\subsubsection{Portage: HPG-MxP}
\label{sec:bench-portage}

We ran HPG-MxP benchmark experiments on Portage using Instinct MI300A
APUs, which integrate CPU cores, CDNA3 GPU tiles, and HBM3 in a single
package and are measured at the APU level. As described in
Section~\ref{sec:frontier-portage-sensors}, this changes the scope of
the on-chip and Cray~PM sensors: APU power now includes combined
CPU+GPU+HBM activity, and two of the four APUs share power rails with
the Cassini NIC.

The HPG-MxP results on Portage are shown in Figure~\ref{fig:portage-hpg-1x2}, which compares the full-precision trace on the left with the mixed-precision trace on the right. The stacked traces expose compute- and Krylov-dominated phases together with lower-power communication segments, again demonstrating phase-level attribution rather than only total-energy accounting. The mixed-precision run compresses the main compute window and modestly lowers average APU power, so the total energy reduction reflects both shorter time-to-solution and a smaller change in instantaneous power. Across 128 independent runs on 128 nodes, the average node energy consumption was 55.2\,KJ for the full-precision configuration and 40.3\,KJ for the mixed-precision configuration, with standard deviations of 2.9\,KJ and 1.9\,KJ, respectively, corresponding to an average reduction of 27\%.
We also find that our derived power from cumulative energy integrated from on-chip APU energy counters agrees well with Cray~PM node-level power once the NIC offsets are applied. This demonstrates that the same $\Delta E / \Delta t$ reconstruction and phase-level attribution workflow transfers from discrete Instinct MI250X GPUs to integrated Instinct MI300A APUs despite their different sensing scopes and packaging.

\subsubsection{Summary of Benchmark Findings}
\label{sec:bench-summary}

Across rocHPL, rocHPL-MxP, and HPG-MxP on Frontier and Portage,
our benchmark studies confirm three main observations.

First, instantaneous power derived from a few milliseconds of energy counters captures phase boundaries and transient behavior that are strongly smoothed in vendor-averaged power metrics while still matching the steady-state levels reported by Cray~PM. This enables robust alignment of power traces with application phases even when those phases are hundreds of milliseconds long and interleaved with communication.

Second, mixed precision primarily improves energy efficiency by shortening time-to-solution rather than dramatically reducing instantaneous power. On Frontier, mixed-precision modes reduced total energy consumption by 79\% for rocHPL-MxP and 31\% for HPG-MxP compared to their full-precision counterparts. Similar savings were observed on Portage at 81\% and 27\%, respectively. These percentages describe the benchmark behavior; the methodological contribution is that the approach separates runtime effects from changes in instantaneous power.

Third, the component-level stacked power views show that accelerator packages dominate node power during compute phases on both systems, with CPUs, memory, and NICs contributing smaller, largely static baselines. The resulting time-aligned traces are directly useful for current systems, for example, to identify short excursions relevant to power capping, scheduling, and library tuning, and they are also friendly to downstream analytics because they preserve per-phase context, device scope, and MPI alignment.

\section{Related Work}
Several prior studies have examined hardware sensors for fine-grained power and energy measurement. At the CPU level, Khan~\cite{10.1145/3177754} characterized the accuracy, temporal granularity, and overhead of Intel’s RAPL interface on Skylake, with extensive follow-on work validating RAPL across workloads and systems for application-level power analysis~\cite{diouri2014scis, dongarra2012cgc}.

For GPUs, Yang~\cite{yang2024accurate} showed that NVIDIA’s \texttt{nvidia-smi} telemetry depends on GPU generation and samples only $\sim$25\% of execution time on recent devices, leading to under-reported energy. They proposed estimation-based methods to reduce error, motivating the inclusion of energy counters in NVML~\cite{nvidia_nvml_api}. 
Jain~\cite{jain2026minos} derived power from AMD energy counters for workload classification and frequency-capping prediction. By contrast, our work compares derived power against on-chip, off-chip, and node-level sensors. Additional work~\cite{10.1007/978-3-032-07612-0_18} studies fine-grained power behavior and power capping on NVIDIA GH200 using \texttt{hwmon} and NVML. 

On AMD GPUs, tools such as OmniStat~\cite{10.1145/3731599.3767564} expose \texttt{rocm-smi} and Cray PM data for monitoring but do not enable fine-grained attribution. Other studies~\cite{10.1145/3757348.3757363} lack sufficient sensor characterization for accurate application-level attribution. FinGraV~\cite{singhania2024finegrain} addresses sub-millisecond kernels on MI300X using execution-time binning, CPU–GPU clock synchronization, and power-profile differentiation to mitigate jitter and averaging, improving peak power estimation.

To our knowledge, this work is among the first to jointly characterize on-chip and off-chip power and energy sensors on AMD MI250X and MI300 systems in Cray EX environments, and to analyze how these characteristics affect fine-grained attribution, including validation of energy-to-power derivation against sensor-reported power.

\section{Conclusion}
\label{sec:conclusion}

Modern accelerator-based supercomputers expose a rich yet complex ecosystem of power and energy sensors. On systems such as Frontier and Portage, developers must navigate on-chip \texttt{rocm-smi} or \texttt{amd-smi} counters, Cray Power Management (PM) node sensors, and asynchronous tracing tools, all of which differ in scope, update rate, and filtering behavior. This paper presents a measurement model and experimental methodology for characterizing these sensors, reconstructing near-instantaneous power from 1~ms energy counters, and aligning heterogeneous sensor streams with application phases at millisecond resolution. 

Using synthetic square-wave workloads, we quantified update intervals, response/recovery behavior, and aliasing for on- and off-chip sensors on AMD MI250X GPUs and MI300A APUs. We found that energy counters are consistent across scopes, and differentiating them yields instantaneous power that matches off-chip Cray PM in steady state while preserving sharp phase boundaries and short transients. In contrast, native on-chip power counters are often heavily filtered (especially MI250X \texttt{rocm-smi} average power), masking short phases and worsening aliasing. Computing power from energy bypasses this filtering and can provide 1 ms measurements, though instrumentation overhead can stretch reads to a few milliseconds, more so on MI300A, where overhead and smoothing push safe attribution toward tens of milliseconds.

We integrated these insights into an open source Score-P/PAPI-based tracing
framework that samples on-chip and Cray PM sensors asynchronously and
attributes power and energy to application regions, processes, and
devices. Applied to rocHPL, rocHPL-MxP, and HPG-MxP on Frontier and Portage, this framework enables phase-level energy attribution, exposes short excursions that are relevant to power-aware tuning on current systems, and yields synchronized traces that are suitable for scalable analytics. For the studied workloads, mixed precision reduces average node energy consumption by up to 81\% across both clusters, primarily through shorter time-to-solution rather than large drops in instantaneous power. Across both systems, stacked component-level power profiles show that GPU/APU packages dominate node consumption during compute phases, with CPUs, memory, and NICs contributing relatively small, stable baselines.

This work advances fine-grained measurement-driven power and energy optimization for exascale systems by characterizing the sensors available on the nodes of Frontier and Portage. Fine-grained, cross-scope attribution is essential for enabling power-aware optimization as future systems operate under tighter power and thermal constraints.

\section{Acknowledgements}
This research used resources from the Oak Ridge Leadership Computing Facility, which is a US Department of Energy (DOE) Office of Science user facility supported under contract DE-AC05-00OR22725. This material is based upon work supported by the U.S. Department of Energy, Office of Science, Office of Advanced Scientific Computing Research under contract number DE-AC05-00OR22725. This research used resources of the Oak Ridge Leadership Computing Facility at the Oak Ridge National Laboratory, which is supported by the Advanced Scientific Computing Research programs in the Office of Science of the U.S. Department of Energy under Contract No. DE-AC05-00OR22725.



\clearpage
\appendix
\label{appendix:A}

\section{Overview of Cray PM Measurement Counters}

The Cray Power Management (PM) subsystem measures electrical input to node
components \emph{upstream} of the final voltage regulators on the motherboard.
These upstream measurements capture power before device-specific step-downs
(e.g., GPU/APU VRMs), which leads to PM readings that systematically exceed
the on-chip power values reported by \texttt{rocm-smi} or vendor interfaces. For Frontier
(EX235a) nodes this difference is typically within 5\%, while on Portage
(EX255a) nodes the difference is approximately 1\% due to the more tightly
integrated AMD Instinct\texttrademark~MI300A package.

Because PM counters operate at the board level, they also capture contributions
from auxiliary components such as Slingshot NICs, node controllers, and
backplane logic. These details affect attribution and are documented below.

\subsection{Mechanical and Electrical Placement of Sawtooth NIC Cards}

Each EX235a and EX255a node includes two \emph{sawtooth} carrier cards, each
hosting two HPE Cassini Slingshot NICs for a total of four NICs per node.

\paragraph{Frontier (EX235a).}
On Frontier, the NICs draw power from the node baseboard rather than from
individual GPUs. Consequently, all NIC power is included in the node-level PM
counters (\texttt{power}, \texttt{energy}) but excluded from the per-GPU PM
counters (\texttt{accel[0--3]}). This design contributes an additional
$\sim$60\,W to the node-level PM totals under idle network conditions.

\paragraph{Portage (EX255a).}
On Portage, the mechanical packaging differs: each of the two sawtooth cards is
mounted directly above APU~0 and APU~2. The majority of the NIC power is
delivered through the associated APU’s 48\,V rail, with additional power flowing
through PCIe~x16 auxiliary connections shared across the paired APUs. As a
result:
\begin{itemize}
    \item \texttt{accel[0]} and \texttt{accel[2]} PM counters include NIC power,
    \item \texttt{accel[1]} and \texttt{accel[3]} are unaffected, and
    \item the node-level PM counters include the aggregate contribution.
\end{itemize}

This asymmetry is unique to the EX255a layout and must be corrected during APU-
level attribution.

\subsection{Derivation of the 30\,W NIC Offset on Portage}

Under network-quiet operating conditions (low Slingshot traffic and no
multi-node communication), we measured the steady-state idle power of:
\begin{enumerate}
    \item APU-only power using \texttt{rocm-smi} energy counters,
    \item per-APU PM counters for all four APUs, and
    \item the node-level PM total.
\end{enumerate}

Across the 50 sampled EX255a nodes, the PM measurements for APU~0 and APU~2
were consistently higher than the corresponding APU-only \texttt{rocm-smi} values by
approximately $30 \pm 2$\,W per sawtooth card. APU~1 and APU~3 did not exhibit
this offset.

To correct for this static NIC power, we adjust the PM-based accelerator
readings as follows:
\[
\begin{aligned}
\texttt{accel[0]}_{\text{corrected}} &= \texttt{accel[0]}_{\text{PM}} - 30\,\mathrm{W}, \\
\texttt{accel[2]}_{\text{corrected}} &= \texttt{accel[2]}_{\text{PM}} - 30\,\mathrm{W}.
\end{aligned}
\]

This adjustment removes only the static NIC contribution; dynamic NIC power
during communication-heavy phases remains fully captured by the PM sensors.

\subsection{Voltage Conversion and Step-Down Differences}

Because the PM counters measure electrical power upstream of final voltage
regulators, they capture:
\begin{itemize}
    \item VRM conversion losses,
    \item board-level distribution losses, and
    \item the combined effect of pre-regulation filtering.
\end{itemize}

By contrast, \texttt{rocm-smi} reports:
\begin{itemize}
    \item GPU/APU package power \emph{after} step-down regulators,
    \item power within the device’s final regulated voltage domain, and
    \item moving-average or filtered instantaneous power values.
\end{itemize}

This difference explains why:
\[
P_{\text{PM}} > P_{\texttt{rocm-smi}}
\]
for both MI250X and MI300A devices in steady-state operation.

\subsection{On-Chip and Off-Chip Sensor Layout on Frontier and Portage Nodes}

\begin{figure*}[htbp]
    \centering
    \begin{subfigure}[b]{0.48\textwidth}
        \centering
        \includegraphics[width=\linewidth]{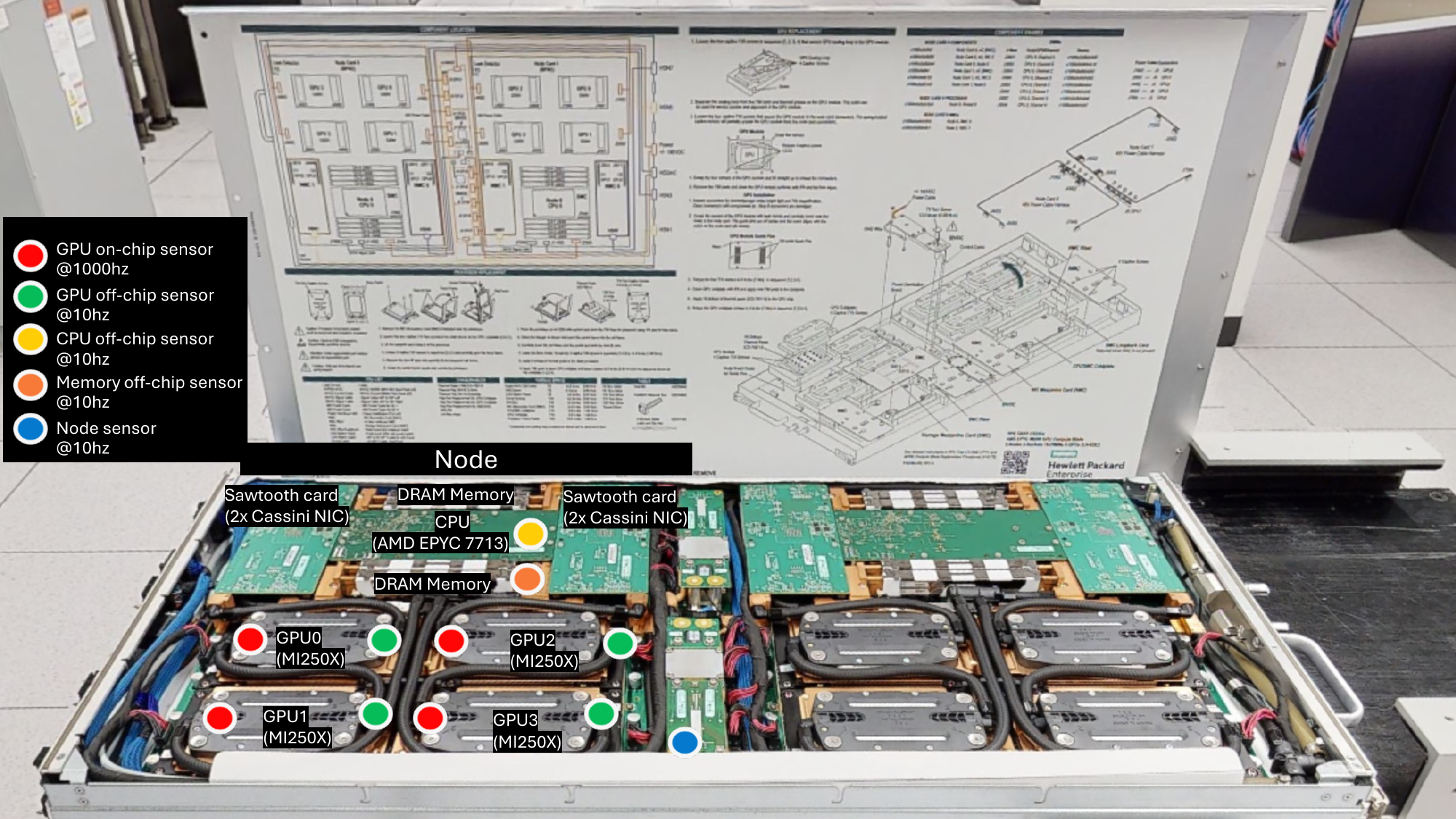}
        \caption{Frontier node (AMD Instinct\texttrademark~MI250X GPUs)}
        \label{fig:node-mi250x}
    \end{subfigure}
    \hfill
    \begin{subfigure}[b]{0.48\textwidth}
        \centering
        \includegraphics[width=\linewidth]{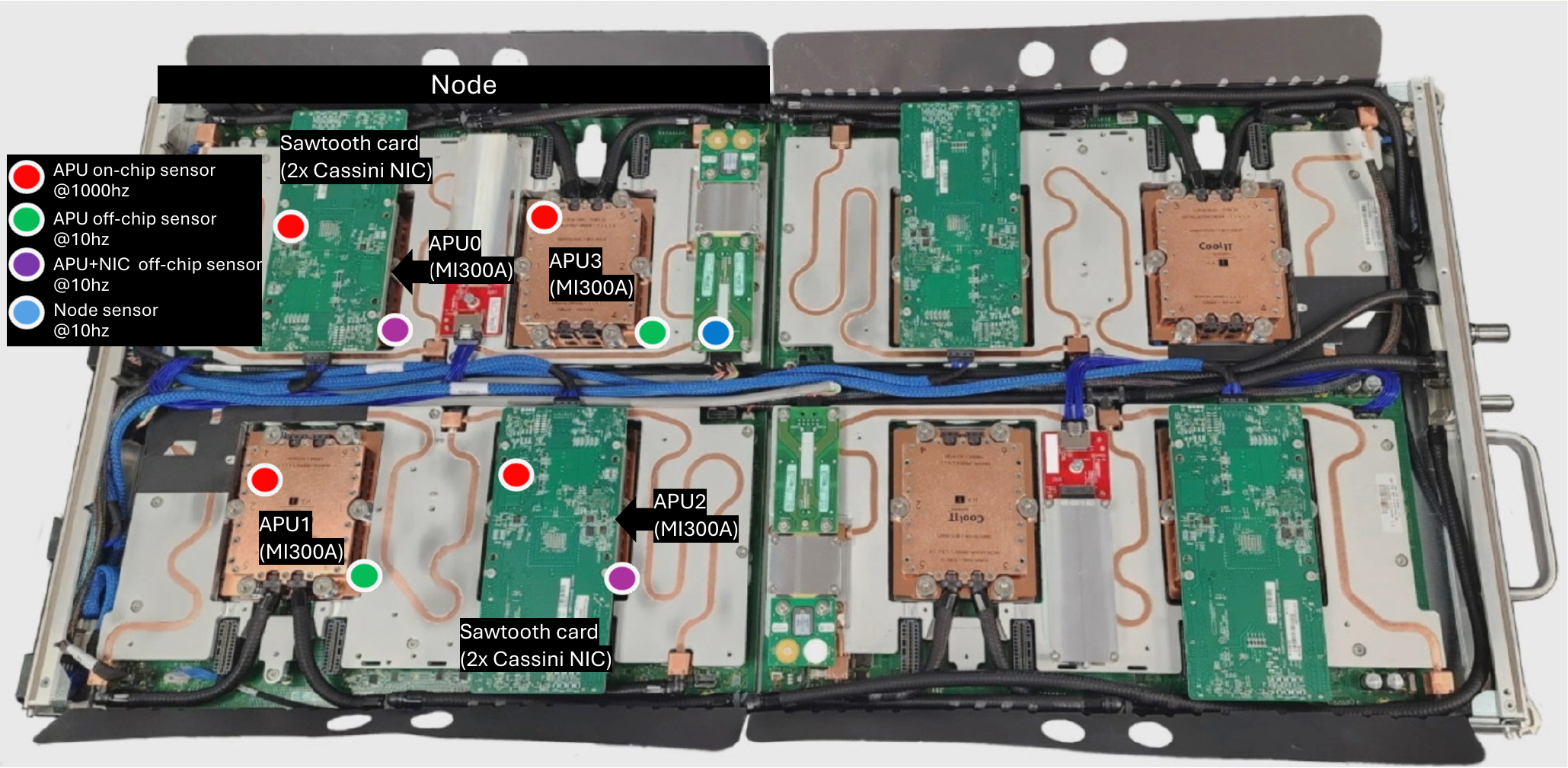}
        \caption{Portage node (AMD Instinct\texttrademark~MI300A APUs)}
        \label{fig:node-mi300a}
    \end{subfigure}
    \caption{Different GPU/APUs power and energy sensor on Frontier and Portage nodes, highlighting on-chip, off-chip, and node-level measurements and their hardware scopes.}
    \label{fig:node-sensors}
\end{figure*}

Figure~\ref{fig:node-sensors} reproduces the annotated hardware
diagrams showing the placement and scope of on-chip and off-chip sensors for
Frontier and Portage nodes. These diagrams highlight the domains measured by
PM counters versus ROCm-SMI.

\subsection{On chip and off chip sensor reference for Frontier and Portage nodes}
\label{app:sensor-counters}

\begin{table}[!htbp]
\centering
\begin{tabular}{|l|l|p{3.6cm}|}
\hline
Counter & Unit & Description \\
\hline
accel[0-3]\_energy & Joules & GPU energy \\
accel[0-3]\_power & Watts & GPU power \\
accel[0-3]\_power\_cap & Watts & GPU power cap, requires admin privileges \\
cpu0\_temp & Degrees C & CPU temperature  \\
cpu\_energy & Joules & CPU energy \\
cpu\_power & Watts & CPU power \\
memory\_energy & Joules & Node memory energy \\
memory\_power & Watts & Node memory power \\
energy & Joules & Energy drawn by node motherboard \\
power & Watts & Power drawn by node motherboard \\
power\_cap & Watts & Node power cap, read only \\
\hline
\end{tabular}
\caption{Cray PM counters on Frontier nodes.}
\label{tab:pm_frontier}
\end{table}

\begin{table}[!htbp]
\centering
\begin{tabular}{|l|l|p{3.6cm}|}
\hline
Counter & Unit & Description \\
\hline
accel[0-3]\_energy & Joules & APU (CPU+GPU) energy \\
accel[0-3]\_power & Watts & APU (CPU+GPU) power \\
accel[0-3]\_power\_cap & Watts & APU power cap, requires admin privileges \\
energy & Joules & Energy drawn by node motherboard \\
power & Watts & Power drawn by node motherboard \\
power\_cap & Watts & Node power cap, read only \\
\hline
\end{tabular}
\caption{Cray PM counters on Portage nodes.}
\label{tab:pm_portage}
\end{table}

\begin{table}[!htpb]
\centering
\begin{tabular}{|l|l|p{3.2cm}|}
\hline
Counter & Unit & Description \\
\hline
energy\_count & Microjoules & Accumulated GPU energy \\
gpu\_clk\_freq\_system & GHz & GPU clock frequency \\
temp\_current:sensor=1 & Millidegrees C & GPU junction temperature \\
power\_average & Microwatts & GPU power moving average \\
busy\_percent & \% & GPU processor usage \\
memory\_busy\_percent & \% & GPU memory usage \\
\hline
\end{tabular}
\caption{\texttt{rocm-smi} metrics on Frontier nodes.}
\label{tab:rocm_frontier}
\end{table}

\begin{table}[!htpb]
\centering
\begin{tabular}{|l|l|p{3.2cm}|}
\hline
Counter & Unit & Description \\
\hline
energy\_count & Microjoules & Accumulated GPU energy \\
gpu\_clk\_freq\_System & GHz & GPU clock frequency \\
temp\_current:sensor=1 & Millidegrees C & GPU junction temperature \\
current\_socket\_power & Microwatts & GPU power \\
busy\_percent & \% & GPU processor usage \\
memory\_busy\_percent & \% & GPU memory usage \\
\hline
\end{tabular}
\caption{\texttt{rocm-smi} metrics on Portage nodes.}
\label{tab:rocm_portage}
\end{table}

\subsection{Aliasing Effect Observed in Frequency Domain}
\begin{figure*}[htbp]
    \centering
    
    \begin{subfigure}{0.48\textwidth}
        \centering
        \includegraphics[width=\linewidth]{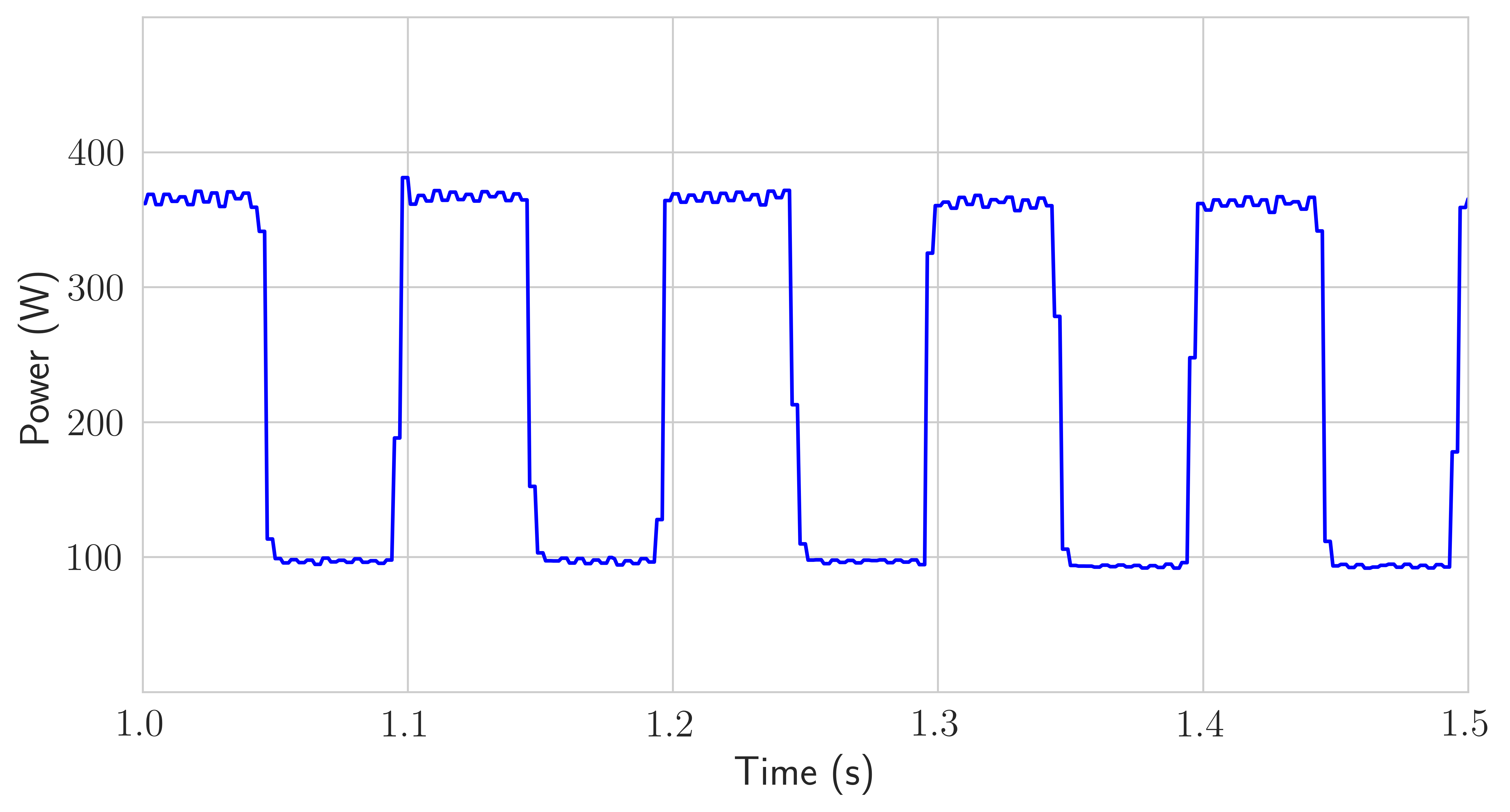}
        \caption{Time Domain: No Aliasing (10 Hz /  100ms period)}
        \label{fig:time_clean}
    \end{subfigure}
    \hfill
    \begin{subfigure}{0.48\textwidth}
        \centering
        \includegraphics[width=\linewidth]{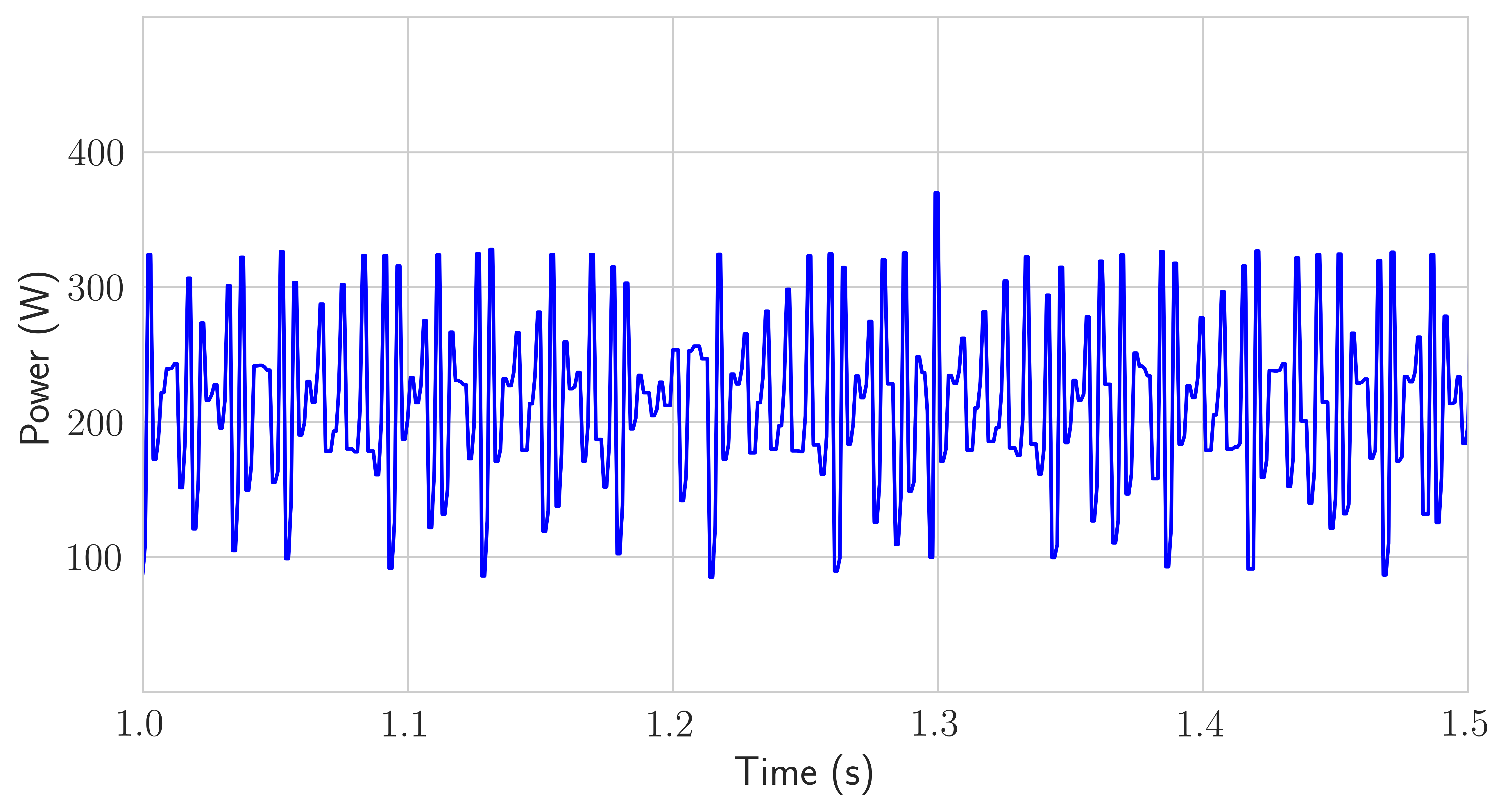}
        \caption{Time Domain: Aliased (250 Hz / 4ms period)}
        \label{fig:time_aliased}
    \end{subfigure}
    
    \vspace{0.5cm} %
    
    \begin{subfigure}{0.48\textwidth}
        \centering
        \includegraphics[width=\linewidth]{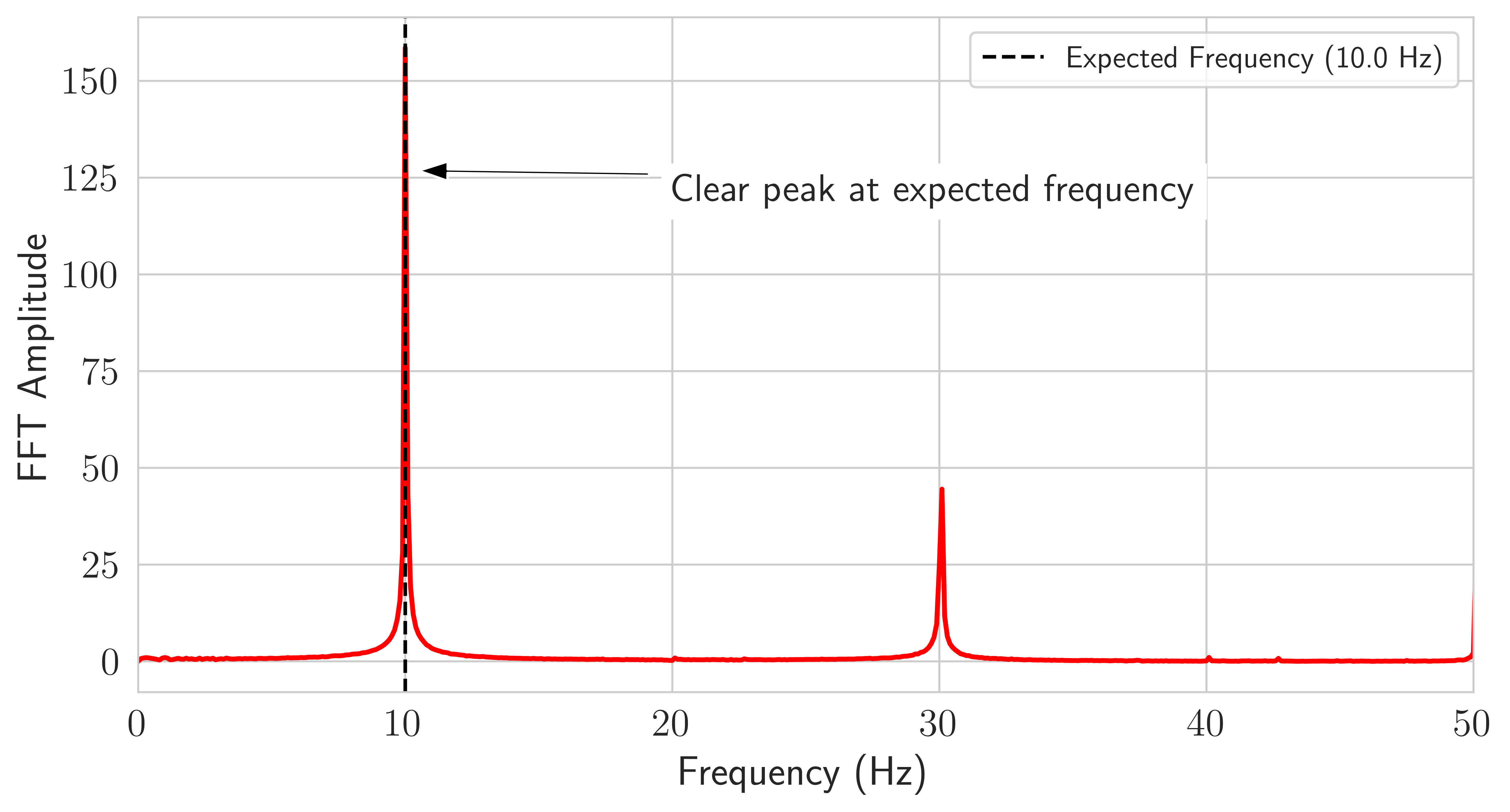}
        \caption{Fourier Spectrum: No Aliasing (10 Hz / 100ms period)}
        \label{fig:freq_clean}
    \end{subfigure}
    \hfill
    \begin{subfigure}{0.48\textwidth}
        \centering
        \includegraphics[width=\linewidth]{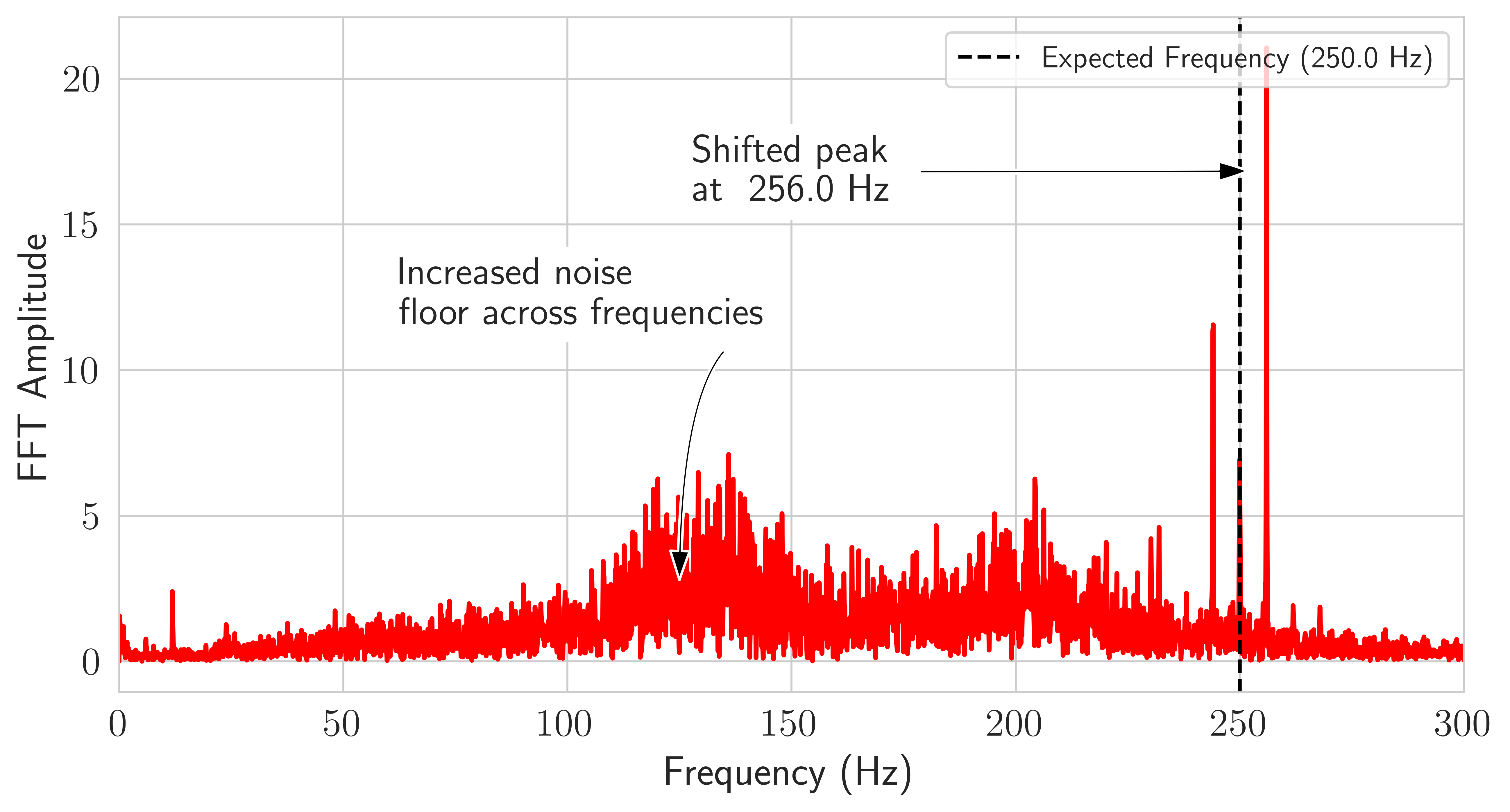}
        \caption{Fourier Spectrum: Aliasing Effect (250 Hz / 4ms period)}
        \label{fig:freq_aliased}
    \end{subfigure}
    
    \caption{Time and frequency domain visualizations for two square-wave workloads measured using derived power from {\tt rocm-smi}'s energy counters on Frontier. Top row shows the time domain observed signals. Bottom row shows the Fourier spectra: (c) A low-frequency workload where the instrumentation's sampling rate outpaces the GPU activity, resulting in clear harmonics at expected frequencies. (d) A higher-frequency workload slightly exceeding the instrumentation's sampling capability, exhibiting a peak shift from the ground-truth period, and an increased floor of noise across all other frequencies.}
    \label{fig:fft-spectrum}
\end{figure*}

To illustrate the impact of the sampling frequency on observed aliasing behavior, we performed FFT operations on two exemplar square wave {\tt rocm-smi} derived power measurement signals using the approach from Sec.~\ref{subsec:deriving_power} on Frontier, a low frequency 10 Hz workload and a high frequency 250 Hz workload.  As can be seen in Fig.~\ref{fig:fft-spectrum}, the sensor is able to accurately capture the behavior of the low frequency workload, showing a peak frequency at 10 Hz. For the  high frequency derived power signal, the time domain plot shows values observed via the {\tt rocm-smi} sensor and the frequency domain plot illustrates the impact of aliasing, with peak frequency shifted from its true value and higher noise frequencies throughout the spectrum. 

\end{document}